\documentclass[12pt]{article}

\usepackage{amsmath}
\usepackage{amssymb}
\usepackage{amsfonts}
\usepackage{graphicx}
\usepackage{slashed} 
 
\numberwithin{equation}{section}

\begin{document}

\begin{center}
{\large \bf{ On the new  LHC resonance }}
\end{center}

\vspace*{2cm}

\begin{center}
{
Paolo Cea~\protect\footnote{Electronic address:
{\tt Paolo.Cea@ba.infn.it}}  \\[0.5cm]
{\em Dipartimento di Fisica, Universit\`a di Bari,  via G. Amendola 173, 70126 Bari, Italy}\\[0.3cm]
{\em  INFN - Sezione di Bari,  via G. Amendola 173, 70126 Bari, Italy} }
\end{center}

\vspace*{1.5cm}

\begin{abstract}
\noindent 
We present an alternative interpretation within the Standard Model of the new LHC resonance at $125 \; GeV$. 
We further elaborate on our previous proposal that  the resonance at 125 GeV could be interpreted as a pseudoscalar 
 meson with quantum number $J^{PC} = 0^{- +}$. We develop a phenomenological approach where  this pseudoscalar  
 mimics the decays of the Standard Model  Higgs boson in the  vector  boson decay channels.
We  propose that  the true Higgs boson should be a heavy resonance with mass of $750 \, GeV$ as argued in 
Ref.~\cite{Cea:2012}. We determine the most relevant decay modes and estimate the partial decay widths and branching ratios.
 We also discuss  briefly  the experimental signatures of  this heavy Higgs boson. Finally, we attempt a  comparison 
of our theoretical expectations  with recent data at $\sqrt{s} =13 \,TeV$ from ATLAS and CMS experiments in the so-called
golden channel. We find that the available experimental data could be consistent with the heavy Higgs scenario. 
\end{abstract}
\newpage

\tableofcontents

\newpage

\section{Introduction}
\label{s-1}
\begin{figure}[t]
\centering
\includegraphics[width=0.90\textwidth,clip]{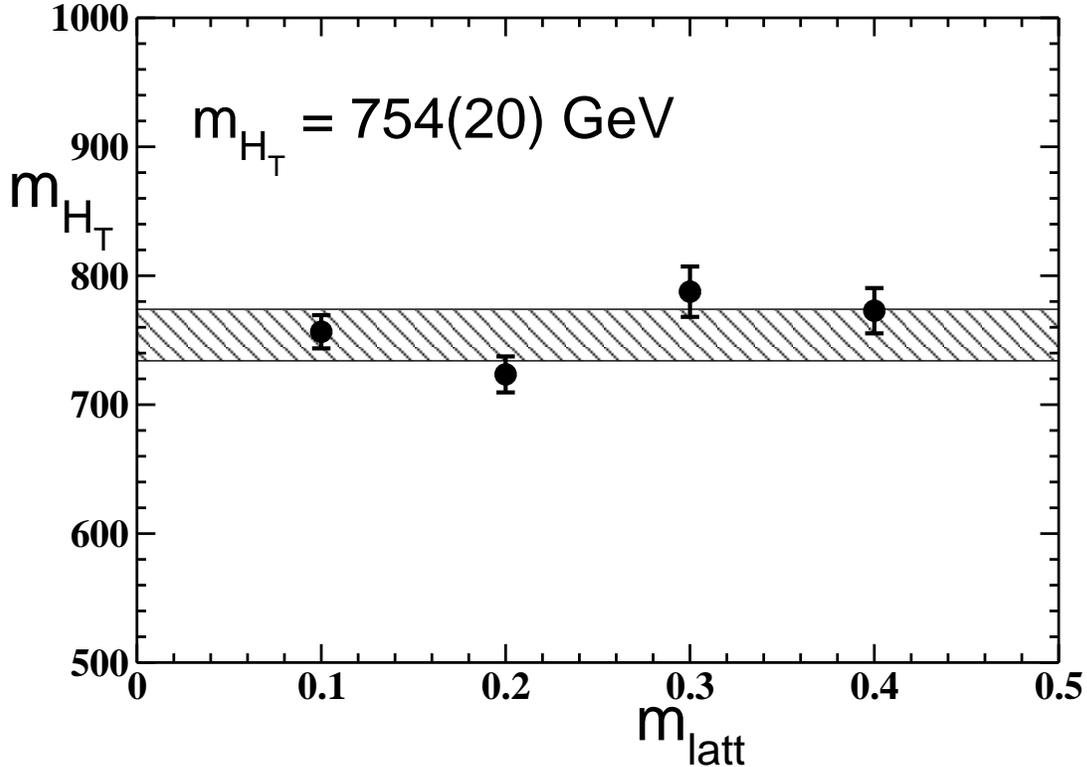}
\caption{\label{Fig-1} Extrapolation to the continuum limit of the $H_T$ Higgs boson mass. The shaded area corresponds to
one standard deviation error in the determination of $m_{H_T}$.  Data  adapted from Fig.~3 of Ref.~\cite{Cea:2012}.}
\end{figure}

A cornerstone of the Standard Model is the mechanism of spontaneous symmetry breaking that is now called the
Brout-Englert-Higgs (BEH) mechanism~\cite{Englert:1964,Higgs:1964,Guralnik:1964,Higgs:1966}.
In fact, the discovery of the so-called Higgs boson  is one of the primary goals of  Large Hadron Collider (LHC) experiments.
The first run of proton-proton collisions at the CERN Large Hadron Collider  has brought the confirmation of the existence of 
a  boson, named H, which resembles, so far, the one which breaks the electroweak symmetry in the Standard Model
 of particle physics~\cite{Aad:2012,Chatrchyan:2012}. 
The combined ATLAS and CMS experiment best estimate of the mass of the H boson is 
$m_H  = 125.09 \,  \pm \, 0.24 \; GeV$~\cite{ATLAS-CMS:2015}.  Results from both LHC experiments, 
as summarized in Refs.~\cite{CMS:2015,ATLAS:2016,ATLAS-CMS:2016},
showed that all measurements of the properties of the new H resonance are consistent with those expected for the
Standard Model Higgs boson. Actually, in the LHC Run 1 the strongest signal significance has been obtained from the
decays of the H boson into two vector bosons, $H \rightarrow V V$ where $V = \gamma, W, Z$. In fact, in these channels
the observed signal significance is above $5 \; \sigma$~\cite{CMS:2015,ATLAS:2016,ATLAS-CMS:2016}. Moreover,
if one introduce the signal strength $\mu$, defined as the ratio of the measured H boson rate to the Standard Model Higgs boson 
prediction, then it turns out that for these decays $\mu \simeq  1$ within the statistical uncertainties.  Nevertheless, these measurements
rely predominantly on studies of the boson decay modes. To establish the mass generation mechanism for fermions as
implemented in the Standard Model, it is of paramount importance to demonstrate the direct coupling of the H resonance to fermions
and the proportionality of its strength to the fermion mass. According to the Standard Model, if one assumes that 
the H resonance at $125 \; GeV$
 is the Higgs boson, then the most common decay of H  should be a transformation into a pair of  
  $bottom-antibottom$  quarks. 
Indeed, the $H \rightarrow b \, \bar{b}$ decay mode is predicted in the Standard Model to have the largest branching ratio. In spite
of this large branching ratio, an inclusive search for  $H \rightarrow b \, \bar{b}$ is not feasible because of the overwhelming background
from multi-jet production. Associated production of  a Higgs boson with vector bosons $W$ or $Z$ offers a viable alternative
notwithstanding a cross section more than an order of magnitude lower than the inclusive production cross section.
In this case a sophisticate statistical analysis is required to fully characterize the background. In general, the number of expected background
events and the associated kinematic distributions are derived from a mixture of data-driven methods and simulations. Indeed,
the shapes of all backgrounds are estimated from Monte Carlo simulations and maximum likelihood fits.
Both the ATLAS and CMS collaborations reported evidences for the  $H \rightarrow b \, \bar{b}$ mode~\cite{ATLAS:2015b,CMS:2014},
albeit with a statistical significance of no more than about $3 \, \sigma$. It should be mentioned, however, that
the ratio of the branching ratios $Br(H \rightarrow b \, \bar{b})/ Br(H \rightarrow Z \, Z)$ displays a deficit of about three standard deviations
relatively to the expected Standard Model value~\footnote{See Table~9 in Ref.~\cite{ATLAS-CMS:2016}.}.
In other words, the $H$ boson seems to decay into a $bottom-antibottom$ pair less frequently than expected. 
Likewise, the experimental evidence for the  $H \rightarrow \tau^{+} \, \tau^{-}$  decay mode has reached  a statistical significance of about 
 $3 \, \sigma$~\cite{ATLAS:2015c,CMS:2014b}. Finally, the spin and $CP$ properties of the $H$ boson can be determined by studying
 the tensor structure of its interactions with the electroweak gauge bosons.  The experimental analyses~\cite{ATLAS:2015d,CMS:2015b}
 rely on  discriminant observables chosen to be sensitive to the spin and parity of the signal. Then, a likelihood function that depends
 on the spin-parity assumption of the signal is constructed. In this way it was possible to compare the Standard Model
hypothesis $J^P = 0^{+}$ to alternative models. The statistic test used to distinguish between two alternative spin-parity
hypotheses was based on the ratio of likelihoods. It turned out that all tested alternative models were excluded with a 
statistical significance of about $3 \, \sigma$~\cite{ATLAS:2015d,CMS:2015b}.  In particular, the Standard Model hypothesis
has been also compared with  an alternative spin-zero pseudoscalar boson  $J^P = 0^{-}$. The pseudoscalar hypothesis
was implemented with effective higher-dimension operators to describe the interactions of the pseudoscalar boson with
the Standard Model vector bosons. \\
To summarize ATLAS and CMS have combined their analyses for production and decay of the $H$ boson. 
Up to now, if one identify the $H$ resonance with the Standard Model Higgs boson, then it turns out that many
results are in agreements with the Standard Model predictions. However, there are some puzzling deviations with respect
to expectations. Aside from the already mentioned deficit in the $bottom-antibottom$ decays, in our opinion the most 
interesting discrepancy manifest itself in the associate  production of the $H$ resonance with a $t \bar{t}$ pair. Indeed,
let us introduce the interaction strength $\mu_{t \bar{t} H}$ defined as the ratio of the observed associate production
cross section to the expected Standard Model value. The observed value of this interaction strength~\cite{ATLAS-CMS:2016}:
\begin{equation}
\label{1.1} 
\mu_{t \bar{t} H}  \;  = \; 
 \left\{ \begin{array}{ll}
 \; \; 2.9 \; ^{+ 1.0}_{- 0.9}  \; \; & CMS  \; \; \;   \; \; Run \, 1
  \\
   \\
  \; \; 1.9  \;  ^{+ 0.8}_{- 0.7}   \; \; &  ATLAS  \;  Run \, 1
\end{array}
    \right.
\end{equation}
deviates from the Standard Model expectations with a  statistical significance of about $2 \, \sigma$. Such deviations,
 if confirmed in the LHC Run 2, may suggest an alternative interpretation of the $H$ resonance. Indeed,
in the LHC Run 2 at $\sqrt{s} =  13 \, TeV$  the associate production of the Higgs boson with  $t \bar{t}$ pairs has
a  cross section which is about four times  larger than in the Run 1.  Remarkably,  both ATLAS~\cite{ATLAS:2016ttH} and
CMS ~\cite{CMS:2016ttH} experiments reported new measurements of the  interaction strength $\mu_{t \bar{t} H}$ 
using LHC collision data at a center of mass energy of $\sqrt{s} = 13 \, TeV$ based on an integrated luminosity of  $13.3 \, fb^{-1}$
 and  $12.9 \, fb^{-1}$ respectively:
\begin{equation}
\label{1.2} 
\mu_{t \bar{t} H}  \;  = \; 
 \left\{ \begin{array}{ll}
 \; \; 1.8 \; ^{+ 0.7}_{- 0.7}   \; \; & ATLAS \;  Run \, 2
  \\
   \\
  \; \; 2.0  \;  ^{+ 0.7}_{- 0.7}   \; \; &  CMS \; \; \; \;  \, Run \, 2
\end{array}
    \right.
\end{equation}
Combining the data from Run 1 and  Run 2,  one obtains:
\begin{equation}
\label{1.3} 
\mu_{t \bar{t} H}  \;  = \; 
 \; \; 2.07 \;  \pm \; 0.38  \; \; ,
\end{equation}
that would imply  a deviation from Standard Model predictions with a statistical significance of about $3 \, \sigma$.
Taking Eq.~(\ref{1.3}) at face value one is led to devastating consequences. In fact, Eq.~(\ref{1.3}) implies that the Higgs coupling to the top
quark is enhanced by a factor $\sim \sqrt{2}$ with respect to the perturbative expectations. Now, observing that
the main production mechanism of the perturbative Higgs boson is by the gluon-gluon fusion processes through top-quark loops, it
follows that the inclusive Higgs production cross section is enhanced by a factor of two thereby reducing the interaction
strengths in the decays into massive vector bosons by the same factor.
Obviously, the observed excess could well be  a statistical fluctuation. Nevertheless, we believe that by now there are compelling 
reasons to look for alternative explanations for the new LHC resonance.
\\
Soon after the evidence of the LHC resonance at  $125 \; GeV$, we proposed~\cite{Cea:2012b} that  the $H$ boson
 could be interpreted as a pseudoscalar meson with quantum number $J^{PC} = 0^{-+}$. The main aim of this paper is to 
 further elaborate the phenomenological approach of Ref.~\cite{Cea:2012b}. In particular, in the first part of the paper
 we will show that  our pseudoscalar meson  could  mimic the decays of the Standard Model  Higgs boson
 in the  vector  boson decay channels, while  the decays into fermions remain  strongly  suppressed. Now, we recall
 that the identification of the $H$ resonance with the Standard Model Higgs boson comes mainly from the decays
 into two vector bosons that reached a statistical significance above $5 \, \sigma$  after the LHC Run 1. Once 
 the pseudoscalar meson could  decay into two vector bosons at the same rate as the Higgs boson, to unravel the true
 nature of the LHC resonance at $125 \, GeV$ one must  rely heavily on the decay modes into two fermions and the $CP$ assignment
 of the resonance.  In this case, however, the reached statistical significance is well below $5 \, \sigma$ and the data display
  a puzzling deficit  in the $b \, \bar{b}$ decay mode. As a consequence the forthcoming  LHC Run 2 data  
  will be crucial to the identification of  the $H$ resonance at $125 \, GeV$ with Standard Model Higgs boson. \\
Adopting the point of view that the new LHC $H$ resonance is not the Higgs boson of the Standard Model,
one faces with the problem of the spontaneous symmetry breaking mechanism and the related scalar Higgs boson. 
 Actually, within the non-perturbative  description of spontaneous symmetry breaking in the Standard Model
it is known that self-interacting scalar fields are subject to the triviality problem~\cite{Fernandez:1992}.
Usually the spontaneous symmetry breaking in the Standard Model is implemented
within the perturbation theory which leads to predict that the Higgs boson mass squared  is proportional to $\lambda \, v^2$,
where $\lambda$ is the renormalized scalar self-coupling and  $v \simeq 246 \; GeV$ is the known weak scale. 
However,  if self-interacting four dimensional scalar field theories are trivial, 
then  $\lambda \rightarrow  0$ when the ultraviolet cutoff  is send to infinity. Strictly speaking, there are  no rigorous proof of triviality.
Nevertheless, there exist several numerical studies  which leave little doubt on the triviality conjecture.  As a consequence, 
within the perturbative approach,  these theories represent just an effective description  valid only up to some cut-off scale. 
On the other hand,  in Ref.~\cite{Cea:2012},  by means of nonperturbative numerical simulations  of the $\lambda \Phi^4$ theory on the lattice,
 it was  enlightened  the scenario where the Higgs boson (denoted as $H_T$ in the following~\footnote{The subscript stands for 
 Trivial or, better, True.}) without self-interaction  could  coexist with spontaneous symmetry breaking.  
Indeed, due to the peculiar rescaling of  the Higgs condensate, the relation between $m_{H_T}$ and the physical $v$ is not the 
same as in perturbation theory.  
According to this picture  the ratio $m_{H_T}/v$  should be a cutoff-independent constant.  Remarkably,  extensive
numerical simulations  showed that the extrapolation to the continuum limit of that ratio leads to a quite sensible result. 
To appreciate this
point, for reader convenience, in Fig.~\ref{Fig-1} we display the numerical results taken from  Fig.~3 of Ref.~\cite{Cea:2012}. 
From  Fig.~\ref{Fig-1} we see that the continuum limit ($m_{latt} \rightarrow 0$) extrapolation of the $H_T$
 Higgs boson mass is consistent  with the intriguing  relation:
\begin{equation}
\label{1.4}
m_{H_T} \; \simeq \;  \pi  \;  v  
\end{equation}
pointing to  a rather massive  $H_T$ boson, $m_{H_T} \; \simeq \; 750$ GeV. 
It is worthwhile to stress that the $H_T$ boson mass almost  exactly matches the  heavy resonance   hinted at the early LHC Run 2 
 in the $\gamma - \gamma$   invariant mass spectrum   by both the ATLAS and CMS
 collaborations~\cite{Aaboud:2016,Khachatryan:2016}. 
Even though in the new LHC Run 2 data the evidence of the $\gamma - \gamma$   
heavy resonance is fading away~\cite{CMS:2016a,ATLAS:2016a,ATLAS:2016b,CMS:2016b},
we shall  identify the Standard Model Higgs boson with the $750 \, GeV$ resonance  $H_T$. Accordingly, in the second part  of the 
present paper  we  further elaborate on the production mechanisms of our  Higgs boson and try to  contrast 
the theoretical expectations with selected available data collected in the LHC Run 2 experiments. \\
The remaining of the paper is organized  as follows. In Sect.~\ref{s-2} we discuss within the Standard Model our proposal for the 
pseudoscalar resonance with mass near $125 \, GeV$. Sect.~\ref{s-3} is devoted to the discussion of the main decay channels of our
pseudoscalar meson within a phenomenological approach. In particular, we explicitly evaluate the partial decay  widths for the decays
in massless vector bosons (Sect.~\ref{s-3.1}), in $Z_0 Z_0$ and  $Z_0 \gamma$  (Sect.~\ref{s-3.2}), in $ W^+ W^-$ 
(Sect.~\ref{s-3.3}), and into fermions (Sect.~\ref{s-3.4}). In  Sect.~\ref{s-3.5}  we estimate the partial decay widths and the
resulting branching ratios.  In Sect.~\ref{s-4} we discuss the production cross section and estimate the interaction strengths. 
In Sect.~\ref{s-5} we briefly illustrate the physics of the heavy $H_T$ Higgs boson. We also estimate the expected production cross 
section. The main decay channels of the $H_T$ Higgs boson are presented in Sect.~\ref{s-5.1}, while 
in Sect.~\ref{s-5.2}  we compare our theoretical expectations with the recent data from LHC in the so-called golden channel.
Finally, Sect.~\ref{s-6}  comprises our concluding remarks. 
\section{The Pseudoscalar Resonance at 125 GeV}
\label{s-2}
The main aim of the present note is to discuss  a possible alternative to the  generally assumed Higgs boson
interpretation of the new LHC resonance at $125 \, GeV$.
In our previous paper we looked for alternative explanations within the Standard Model physics. 
In fact, in Ref.~\cite{Cea:2012b}  we already suggested  that the new resonance could be
interpreted as a pseudoscalar  meson with quantum number $J^{PC} = 0^{- +}$. 
The most natural pseudoscalar candidate within the Standard Model
is a $q \bar{q}$ bound state with L=S=0. Given the large mass of the new resonance we focused on 
the pseudoscalar $t \bar{t}$ that, for obviously reasons,   will be referred  to as $\eta_{t}$. 
 Since the top quark mass is very large:
\begin{equation}
\label{2.1}
m_t \;  \simeq  \;  173 \;  GeV \; ,
\end{equation}
to estimate the mass of the pseudoscalar meson  $\eta_{t}$, we may safely employ the non-relativistic potential model.
Quarkonium potential models typically take the form of a Schr\"odinger like equation:
\begin{equation}
\label{2.2}
\big [ \, T \; + \;  V \, \big ] \, \Psi	\;  =   \;  m \; \Psi
\end{equation}
where $T$ represents the kinetic energy term and $V$ the potential energy term. Equation~(\ref{2.2}) arises from the 
Bethe-Salpeter equation by replacing the full interaction by an instantaneous local potential.
The $quark - antiquark$  potential is typically motivated by the properties expected from QCD.
Assuming that at short distances  V(r) behaves according to perturbative QCD, then the contribution arising from
 one-gluon-exchange leads to the Coulomb like potential. At large distances the one-gluon-exchange is no longer a good representation
 of the $quark - antiquark$  potential.  The qualitative picture is that the chromoelectric lines of force bunch together
  into a flux tube which leads to a distance independent force or linearly rising confining potential. A  potential
 widely used to describe $c \bar{c}$ and $b \bar{b}$ quarkonium states is  the  so-called Cornell potential~\cite{Eichten:1980}:
\begin{equation}
\label{2.3}
V_C  = -\; \frac{4}{3} \, \frac{\alpha_c }{r} \; + \; \sigma \, r  \; , \; \alpha_c \, \simeq \, 0.40 \; , \; \sigma  \simeq \, 0.18 \; GeV^2  \; .
\end{equation}
The empirical coefficient of the short-distance Coulomb potential $\alpha_c  \simeq  0.40$ in the Cornell potential
is much larger than the perturbative QCD expectations. This leads to an overestimate of the $q \bar{q}$ bound states
for very massive quarks.  Alternatively, one may adopt the  Richardson potential~\cite{Richardson:1979} that incorporates both the
QCD asymptotically free short-distance behavior: 
\begin{equation}
\label{2.4}
\lim_{r \rightarrow 0}  \;  V(r) \; \sim \;  - \;   \frac{4}{3} \; \frac{1 }{r \ln r}  
\end{equation}
and the linear confinement potential at large distances. However, long time ago in Ref.~\cite{Cea:1986} it was showed that 
if one assumes that  $V(r)$ behaves according to Eq.~(\ref{2.4}) for $r \rightarrow 0$, then  the
coupling the   $q \bar{q}$ S-wave vector mesons to the electromagnetic current diverges due to the singular behavior of
the wavefunction near the origin. This should imply, for instance, that the ratio $R_{e^+e^-}$ is divergent.  This divergence
is, in fact, an artifact of the instantaneous potential approximation that is not reliable for small enough lengths.
Actually, the authors of  Ref.~\cite{Cea:1986} found that  these spurious divergences could be removed if one assume
a constant potential for scales smaller than some high-energy reference scale $\sim 1/M$. In this way one recover 
the duality between bound states and asymptotically free quarks  leading to canonical results for two-point
spectral functions of vector and axial-vector currents~\cite{Cea:1986}. Accordingly, we may adopt the  Cornell potential 
Eq.~(\ref{2.3}) where now:
\begin{equation}
\label{2.5}
 \alpha_c \;  \simeq \;  \alpha_s(M) \; \simeq \; 0.10 \; \; \; , \; \; \; M \; \simeq \; 10^2 \; GeV \; ,
\end{equation}
with $ \alpha_s(M)$ the coupling constant of strong interactions at the scale $M$.
For our purposes it is enough to reach  a qualitative estimate of the low-lying $L=S=0$ bound state. 
Since the contribution of the linearly rising confining potential can be safely neglected due to the very large top mass, 
we  obtain at once  the wave function of the low-lying $L=S=0$ bound state:
\begin{equation}
\label{2.6}
\Psi_{00}(r) \; \simeq \;  \frac{1}{ ( \pi \;  a_0^3)^{\frac{1}{2}}} \;  \exp(-\frac{r}{a_0} )  \; \; ,
\end{equation}
where $a_0$ is the Bohr radius:
\begin{equation}
\label{2.7}
a_0 \; =  \;  \frac{ 3}{2 \; m_t \; \alpha_c}    \; \; .
\end{equation}
We may, then, easily estimate the pseudoscalar mass as follows:
\begin{equation}
\label{2.8}
m_{\eta_{t}}  \; \simeq  \; 2 \; m_t \; -\; \frac{4}{3} \; \frac{\alpha_c }{a_0} \;  \simeq  \;  344 \; GeV  \; .
\end{equation}
Even though our analysis has been somewhat qualitative, it is evident that the pseudoscalar $\eta_{t}$ meson is too 
heavy to be identified with the new LHC resonance. To overcome this problem we must admit that 
the $\eta_{t}$ meson can have sizable mixing with a much more  lighter   pseudoscalar meson. 
In this regard, we observe that  the self-coupling of gluons in QCD suggests that additional
mesons made of bound gluons (glueballs) may exist. In fact, lattice calculations, flux tube and constituent glue models agree that 
the lights glueballs  have quantum number $J^{PC} = 0^{++}, 2^{++}$  (for a recent review see 
Refs.~\cite{Klempt:2007,Crede:2009}). Moreover, there is a general
agreements on the existence of pseudoscalar states with $J^{PC} = 0^{-+}$  above $2 \, GeV$. 
In the following we will indicate the lowest glueball pseudoscalar state with $\eta_{g}$ and follow the lattice calculations 
for the mass of the lowest pseudoscalar glueball  to set the value~\cite{Crede:2009}: 
\begin{equation}
\label{2.9}
m_{\eta_{g}} \;  \simeq  \;  2.6 \; GeV  \; .
\end{equation}
We see, then, that the pseudoscalar $\eta_{t}$ meson can also mix with the pseudoscalar  $\eta_{g}$ meson through color singlet gluon 
intermediate states. In this case there are no reasons to restrict the intermediate states to two gluons. In fact, if more gluons are involved then 
one gets large effective couplings due to the small typical momentum going into each one making the theory strongly 
coupled~\cite{Appelquist:1975}. If this is the case, then the large top mass gives rise to a sizable mixing amplitude. 
We shall proceed as the authors of  Ref.~\cite{DeRujula:1975} did for the  mesons $\eta$ and $\eta'$. In fact, 
 if we assume that the annihilation process  contribute the flavor independent amount A, we obtain the following mass matrix:
\begin{equation}
\label{2.10}
\cal{M} \; = \;  \left(
\begin{array} {cc}
m_{\eta_g} \; + \; A & A \\
A & m_{\eta_t}  \; + \; A
\end{array}
\right) \; \;.
\end{equation}
The mass matrix can be easily diagonalized by writing the physical mass eigenstates as:
\begin{eqnarray}
\label{2.11}
 \eta_{gt} \;  & = &    \eta_g \; \cos  \theta  \; \; - \; \; \eta_t  \; \sin \theta  
 \nonumber   \\
  \\
 \eta'_{gt} \;  & = &    \eta_g \; \sin \theta  \; \; + \; \; \eta_t  \; \cos \theta  \;
 \nonumber
\end{eqnarray}
where $\theta$ is the mixing angle. Inverting Eq.~(\ref{2.11}) leads to:
\begin{eqnarray}
\label{2.12}
 \eta_{g} \;  & = &    \eta_{gt} \; \cos \theta  \; \; + \; \; \eta'_{gt}  \; \sin \theta  
 \nonumber   \\
  \\
 \eta_{t} \;  & = &    \eta'_{gt} \; \cos \theta  \; \; - \; \; \eta_{gt}  \; \sin \theta   \; \; .
 \nonumber
\end{eqnarray}
We denote with $\eta_{gt}$ the state with lowest mass eigenvalue. Moreover we impose that: 
\begin{equation}
\label{2.13}
m_{\eta_{gt}} \;  \simeq  \;  125 \; GeV \; .
\end{equation}
Then, a standard calculation gives:
\begin{equation}
\label{2.14}
m_{\eta'_{gt}} \;  \simeq  \;  776 \; GeV  \;  \; \; , \; \; \; \theta \; \simeq \; 29 \,^\circ  \; \; .
\end{equation}
Since the mass eigenstate $\eta'_{gt}$ lies well above the $t \bar{t}$ threshold, it hardly can be detected as a hadronic resonance. 
On the other hand, the eigenstate $\eta_{gt}$ could be a serious candidate for the new resonance detected at LHC. 
To corroborate these expectations  we need to estimate the total width and the decay channels of the pseudoscalar meson $\eta_{gt}$.  
\section{Decay Channels}
\label{s-3}
In order to determine the decay rates of the pseudoscalar meson $\eta_{gt}$ we must take care of the fact that it is
a mixture of the pseudoscalar glueball $\eta_g$ and the pseudoscalar $t \bar{t}$ bound state $\eta_t$.
In general,  the decay width depends on the wave function. Since $m_t \gg m_{\eta_g}$, from
Eqs.~({\ref{2.6}) and (\ref{2.7})  we see that the glueball contributions to the decays can be safely neglected.
 Thus, for the decay amplitude of  $\eta_{gt}$ into a generic final state $F$ we can write:
\begin{equation}
\label{3.1}
\mathcal{A}  \big ( \eta_{gt} \rightarrow  F \big )  \; \simeq  \; 
\mathcal{A}  \big ( \eta_{t} \rightarrow F \big )  \; \sin \theta  \; .
\end{equation}
To estimate the contribution of the  $\eta_{t}$  component to the  decay width we may use the well 
known heavy quarkonium model~\cite{Novikov:1978,Close:1979} where the decay amplitudes depend on the bound-state wave
function at the origin. \\
Obviously, the most important decay mode is the single-quark decay, leaving the other quark as a spectator. In fact, since the $t$ quark
is very heavy this single quark decay becomes the dominant mode of t-quarkonium states, hiding all the other modes. However, observing
that the mass of our pseudoscalar $m_{\eta_{tg}} \simeq 125 \, GeV$ is smaller than the $t$ quark mass this decay is forbidden. 
Thus we are left with the decays into two-body final states.
Naively, we expect that  the main decay channels  are given by the decay of  $\eta_{t}$ into ordinary hadrons that,
however, are suppressed by the OZI rule. Therefore, it should turn out that the pseudoscalar meson $\eta_{tg}$ is rather narrow. 
Indeed, the pseudoscalar meson decays involve $t\bar{t}$ quarks annihilation. Due to the very high top quark mass the 
annihilation of a $t\bar{t}$ pair is a short distance process that can be, generally, described by a small effective coupling constant.
Therefore one can safely use perturbation theory to find the corresponding transition amplitude. To evaluate the decay rates of
the pseudoscalar meson  $\eta_{tg}$  into two-body final states within the Standard Model~\footnote{In the following we use the 
convention of Ref.~\cite{Cheng:1982}.} we shall follow the method developed in Refs.~\cite{Kuhn:1979,Guberina:1980,Kuhn:1981}.
According to these authors the amplitude for the decay of a given $quark-antiquark$ bound state into any final state is written
as the convolution of the scattering matrix element with the relevant wavefunction. We are interested in the case of $L=S=0$
$quark-antiquark$ bound state with wave function $\Psi_{00}(r)$. Let $P^{\mu}$ be the four-momentum of the bound state, then
the  decay amplitude can be written as~\cite{Barger:1987}:
\begin{equation}
\label{3.2}
\mathcal{A} \big ( \eta  \rightarrow  F \big )  \; \simeq  \; \frac{1}{2} \; \sqrt{ \frac{3}{M_{\eta}}} \; 
\Psi_{00}(0) \; Tr \Bigg \{ \mathcal{O}_F \,  \gamma_5 \, (- \slashed{P} + M_{\eta})  \Bigg \}  \; ,
\end{equation}
where  the factor $\sqrt{3}$ is due to color. $ \mathcal{O}_F$ is obtained from the Feynman-diagram amplitude for a $quark$
 with momentum   $\frac{P^{\mu}}{2} + q^{\mu}$ 
and an $antiquark$ with momentum  $\frac{P^{\mu}}{2} - q^{\mu}$ to scatter into the final state $F$ by removing the spinor factors.
For comparison with the literature, we recall that the wavefunction  $\Psi_{00}(r)$ is related to the radial wavefunction by:
\begin{equation}
\label{3.3}
\Psi_{00}(r) \; = \; \frac{1}{\sqrt{4 \pi}} \; R_0(r) \;    \; .
\end{equation}
\subsection{$\eta_{tg} \rightarrow \gamma \gamma ,  g g $}
\label{s-3.1}
\begin{figure}
\vspace{-0.5cm}
\begin{center}
\includegraphics[width=1.0\textwidth,clip]{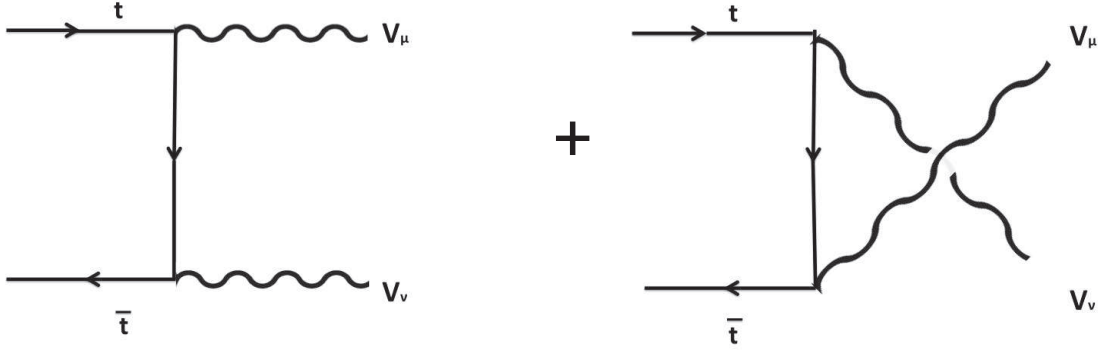}
\caption{\label{Fig-2}  Feynman diagrams contributing to the decay of the pseudoscalar $t\bar{t}$ meson into two
 vector bosons $V_{\mu}$  $V_{\nu}$  ($V V = \gamma \gamma, g g, Z_0 Z_0, Z_0 \gamma$). }
\end{center}
\end{figure}
Our aim is to estimate  the decay  widths  of  the pseudoscalar meson  to  two-body  final  states  within the Standard Model.
 Let us consider, firstly,  the decay into two photons.  Actually, the widths for the decays to two massless vector states are well 
 known in literature~\cite{Novikov:1978,Close:1979}. Indeed,  the relevant Feynman diagram are shown in Fig.~\ref{Fig-2}. 
 A standard calculation gives:
\begin{equation}
\label{3.4}
Tr \bigg \{ \mathcal{O}_{\gamma\gamma} \,  \gamma_5 \, (- \slashed{P} + M_{\eta})  \bigg \}  \; \simeq \; 16 \pi \alpha  \,  Q^2_t \bigg \{
\frac{1}{2 (\frac{P}{2} \cdot k_1)} \;  \epsilon_{\mu\nu\rho\sigma}  \; \varepsilon_1^{\mu} \, \varepsilon_2^{\nu} \, k_1^{\rho} \, 
P^{\sigma} \; + \;  (1 \rightarrow 2) \bigg \} \; ,
\end{equation}
where $\varepsilon_1^{\mu} = \varepsilon^{\mu}(k_1,\lambda_1) $,  $\varepsilon_2^{\nu} = \varepsilon^{\nu}(k_2,\lambda_2) $,  
and $Q_t = \frac{2}{3}$ is the top quark electric charge. Therefore, according to our previous discussion,  we have:
\begin{equation}
\label{3.5}
\mathcal{A} \big ( \eta_{gt}  \rightarrow  \gamma \gamma \big )  \, \simeq  \,  8 \pi \alpha  \,  Q^2_t  \; \sqrt{ \frac{3}{M_{\eta_{gt}}}} \; 
\Psi_{00}(0) \, \sin \theta \;  \bigg \{
\frac{1}{2 (\frac{P}{2} \cdot k_1)} \;  \epsilon_{\mu\nu\rho\sigma}  \; \varepsilon_1^{\mu} \, \varepsilon_2^{\nu} \, k_1^{\rho} \, 
P^{\sigma} \; + \;  (1 \rightarrow 2) \bigg \} \; .
\end{equation}
Aside from the $t \bar{t}$ annihilation contribution we need to take care of transition amplitude due to  quantum anomalies. Indeed,
it is known since long time that the trace and chiral anomalies~\cite{Treiman:1985,Bertlmann:1996}
imply an effective coupling of  pseudoscalar mesons to gauge vector fields. For the electromagnetic field this anomalous coupling
can be obtained from an effective Lagrangian that, following Ref.~\cite{Chanowitz:1984} is written as:
\begin{equation}
\label{3.6}
\mathcal{L}^{eff}_{\gamma \gamma}(x)  \; \simeq \; \frac{ a_{\gamma \gamma}}{\sqrt{6 \pi}}  \;
\frac{\alpha}{f_{\eta_{gt}}} \;  F_{\mu \nu}(x) \,  \widetilde{F}^{\mu \nu}(x) \; \phi_{\eta_{gt}}(x)  
\end{equation}
where $f_{\eta_{gt}}$  is the analogous of the pion decay constant $f_\pi$~\cite{Scadron:1981}.  In Eq.~(\ref{3.6})  $F_{\mu \nu}$
is the electromagnetic field strength tensor,   $\widetilde{F}_{\mu \nu} = \frac{1}{2}   \epsilon_{\mu\nu\rho\sigma}  F^{\rho \sigma}$,
and  $\phi_{\eta_{gt}}$ is a (pseudo-)scalar interpolating quantum field. For low-mass
pseudoscalar mesons, PCAC and low-energy theorems allow to determine the parameter  $a_{\gamma \gamma}$. However,
in the present case since both the pseudoscalar glueball $\eta_g$ and the $t\bar{t}$ pseudoscalar are involved we are not in the position
to offer a reliable estimate of  $a_{\gamma \gamma}$. Therefore, within our phenomenological approach we leave that parameter
as  free. Actually, the unknown parameter is the ratio $a_{\gamma \gamma}/f_{\eta_{gt}}$. In the following, for definiteness we
assume:
\begin{equation}
\label{3.7}
f_{\eta_{gt}} \; \simeq \; 1.0 \; GeV \; 
\end{equation}
while  $a_{\gamma \gamma}$ will be a dimensionless parameter. The effective Lagrangian  Eq.~(\ref{3.6}) gives rise to an additional
transition amplitude:
\begin{equation}
\label{3.8}
\widetilde{\mathcal{A}} \big ( \eta_{gt}  \rightarrow  \gamma \gamma \big )  \, \simeq  \; - \,   \frac{ a_{\gamma \gamma}}{\sqrt{6 \pi}}
\frac{\alpha}{f_{\eta_{gt}}}  \frac{1}{2 }  \;   \epsilon_{\mu\nu\rho\sigma} 
\big [ k_1^\mu \varepsilon_1^{\nu} \, -  \, k_1^\nu \varepsilon_1^{\mu}  \big ] \,
\big [ k_2^\rho \varepsilon_2^{\sigma} \, - \, k_2^\sigma \varepsilon_2^{\rho}  \big ]  \; .
\end{equation}
Whereupon, the partial decay width is:
\begin{equation}
\label{3.9}
\Gamma \big ( \eta_{gt}  \rightarrow  \gamma \gamma \big )  \; =  \; \frac{1}{2!} \; \frac{1}{16 \pi} \; \frac{1}{m_{\eta_{gt}}} \;
\overline{ \bigg |  \mathcal{A} \big ( \eta_{gt}  \rightarrow  \gamma \gamma \big ) \, + \, 
\widetilde{\mathcal{A}} \big ( \eta_{gt}  \rightarrow  \gamma \gamma \big ) \bigg |^2} \; ,
\end{equation}
where:
\begin{equation}
\label{3.10}
\overline{ \bigg |  \mathcal{A} \big ( \eta_{gt}  \rightarrow  \gamma \gamma \big ) \, + \, 
\widetilde{\mathcal{A}} \big ( \eta_{gt}  \rightarrow  \gamma \gamma \big ) \bigg |^2} \;   = \;
\sum_{\lambda_1,\lambda_2} 
 \bigg |  \mathcal{A} \big ( \eta_{gt}  \rightarrow  \gamma \gamma \big ) \, + \, 
\widetilde{\mathcal{A}} \big ( \eta_{gt}  \rightarrow  \gamma \gamma \big ) \bigg |^2 \; .
\end{equation}
To evaluate the polarization average, we put the top quark on the mass shell and neglect the binding energy so that 
$m_{\eta_t} \simeq 2 m_t$. This ensures  the absence of  spurious kinematical threshold  singularities. After that, to
extrapolate to $m_{\eta_{gt}}$ we assumed a dipolar form factor. After some algebra we find:
\begin{eqnarray}
\label{3.11}
\Gamma \big ( \eta_{gt}  \rightarrow  \gamma \gamma \big )  \; \simeq  \;  \frac{\alpha^2}{32 \pi} \; m_{\eta_{gt}} \;
\bigg \{  1536 \, \pi^2 \, Q^4_t  \; \frac{|\Psi_{00}(0)|^2  }{m^3_{\eta_{gt}}} \; \sin^2 \theta \; 
\\  \nonumber
 \; -  \; \frac{ 64 \pi}{\sqrt{2 \pi}} \; Q^2_t  \;   \frac{ a_{\gamma \gamma}}{f_{\eta_{gt}}} \;  \frac{\Psi_{00}(0)  }{\sqrt{m_{\eta_{gt}}}} 
 \; \sin \theta  \; + \;  \frac{ a^2_{\gamma \gamma}}{3 \pi} \;    \frac{m^2_{\eta_{gt}}}{ f^2_{\eta_{gt}} }
 \bigg \} \; .
\end{eqnarray}
Similarly we may calculate the width for the decay into two gluons. As concern the annihilation term, taking into account that the two
gluons must be in a color singlet state, the transition rate for $\eta_{gt} \rightarrow gg$ can be obtained from the rate in two photons
with the replacement $\alpha^2 Q^4_t \rightarrow \frac{2}{9} \alpha_s^2$~\cite{Novikov:1978,Close:1979}.
Even in the present case we have an anomalous coupling to  the gluon fields that  
can be accounted for  by the  effective Lagrangian~\cite{Chanowitz:1984}:
\begin{equation}
\label{3.12}
\mathcal{L}^{eff}_{g g }(x)  \; \simeq \; \frac{ a_{ g g }}{\sqrt{6 \pi}}  \;
\frac{\alpha_s}{f_{\eta_{gt}}} \;  G^a_{\mu \nu}(x) \,  \widetilde{G}_a^{\mu \nu}(x) \; \phi_{\eta_{gt}}(x)  
\end{equation}
The decay width can be obtained by means of calculations very similar to the previous ones. So that, here, we merely present
the final result:
\begin{eqnarray}
\label{3.13}
\Gamma \big ( \eta_{gt}  \rightarrow  g g  \big )  \; \simeq  \;  \frac{\alpha_s^2}{32 \pi} \; m_{\eta_{gt}} \;
\bigg \{  1024 \, \pi^2   \; \frac{|\Psi_{00}(0)|^2  }{m^3_{\eta_{gt}}} \; \sin^2 \theta \; 
\\  \nonumber
 \; -  \; \frac{ 64 \sqrt{\pi}}{3} \;    \frac{ a_{g g}}{f_{\eta_{gt}}} \;  \frac{\Psi_{00}(0)  }{\sqrt{m_{\eta_{gt}}}} 
 \; \sin \theta  \; + \;  \frac{ a^2_{g g }}{3 \pi} \;    \frac{m^2_{\eta_{gt}}}{ f^2_{\eta_{gt}} }
 \bigg \} \; .
\end{eqnarray}
\subsection{$\eta_{tg} \rightarrow Z_0 Z_0 , Z_0 \gamma $}
\label{s-3.2}
Let us, now, focus on the decays into massive gauge vector bosons. In the present Section we consider the decays into 
$Z_0 Z_0$ and $Z_0 \gamma$, while the decay into two $W$ vector bosons will be discussed in the next Subsection. \\ 
The annihilation contribution to the decay $\eta_{gt} \rightarrow Z_0 Z_0$ can be obtained from the Feynman diagrams
in Fig.~\ref{Fig-2}.  The corresponding transition amplitude is given by:
\begin{eqnarray}
\label{3.14}
\mathcal{A} \big ( \eta_{gt}  \rightarrow  Z_0 Z_0 \big )  \, \simeq  \,  \frac{ \pi}{2} \;  \alpha \;  
\frac{1 + (1 - \frac{8}{3} \sin^2 \theta_W)^2}{\sin^2 \theta_W \cos^2 \theta_W} \;
  \sqrt{ \frac{3}{M_{\eta_{gt}}}} \;  \Psi_{00}(0) \, \sin \theta \;  \times
\\ \nonumber
\bigg \{
\frac{1}{(\frac{P}{2} - k_1)^2 - m^2_t} \;  \epsilon_{\mu\nu\rho\sigma}  \; \varepsilon_2^{\mu} \, \varepsilon_1^{\nu} \,
P^{\rho}   k_1^{\sigma} \,   \; + \;  (1 \rightarrow 2) \bigg \} \; .
\end{eqnarray}
In the same manner we find the following annihilation amplitude for the decay into $Z_0 \gamma$:
\begin{eqnarray}
\label{3.15}
\mathcal{A} \big ( \eta_{gt}  \rightarrow  Z_0 \gamma \big )  \, \simeq  \; 2  \pi \;  \alpha \; Q_t \;  
 \frac{1 - \frac{8}{3} \sin^2 \theta_W}{\sin \theta_W \cos \theta_W} \;
  \sqrt{ \frac{3}{M_{\eta_{gt}}}} \;  \Psi_{00}(0) \, \sin \theta \;  \times
\\ \nonumber
\bigg \{
\frac{1}{(\frac{P}{2} - k_2)^2 - m^2_t} \;  \epsilon_{\mu\nu\rho\sigma}  \; \varepsilon_2^{\mu} \, \varepsilon_1^{\nu} \,
  k_1^{\rho} \, P^{\sigma}   \; + \;  (1 \rightarrow 2) \bigg \} \; .
\end{eqnarray}
The quantum anomalies are present also for the $W$ and $Z$ couplings. Here, however, the situation is more involved since,
in general, the anomalies affect differently the axial and vector coupling to the massive boson vectors. One could take care
of this by introducing two more anomalous terms with strengths $a_{ZZ}$ and $a_{WW}$:
\begin{equation}
\label{3.16}
\mathcal{L}^{eff}_{Z Z }(x)  \; \simeq \; \frac{ a_{ Z Z }}{\sqrt{6 \pi} f_{\eta_{gt}} }  \;
\frac{\alpha}{\sin^{2} \theta_W  \cos^{2} \theta_W} \;  Z_{\mu \nu}(x) \,  \widetilde{Z}^{\mu \nu}(x) \; \phi_{\eta_{gt}}(x)  \; ,
\end{equation}
\begin{equation}
\label{3.17}
\mathcal{L}^{eff}_{W W }(x)  \; \simeq \; \frac{ a_{ W W }}{\sqrt{6 \pi} f_{\eta_{gt}} }  \;
\frac{\alpha}{\sin^2 \theta_W} \;  W^{+}_{\mu \nu}(x) \,  \widetilde{W}^{- \, \mu \nu}(x) \; \phi_{\eta_{gt}}(x)  \; 
+ \; h. c. \; .
\end{equation}
After some standard calculations we find the following decay widths:
\begin{eqnarray}
\label{3.18}
\Gamma \big ( \eta_{gt}  \rightarrow Z_0 Z_0  \big )  \; \simeq  \;  \frac{1}{16 \pi} \; \frac{\alpha^2}{\sin^{4} \theta_W  \cos^{4} \theta_W} 
\;  m_{\eta_{gt}} \;  \times
\\ \nonumber
\bigg \{24 \, \pi^2   \; \bigg [\frac{1 + (1 - \frac{8}{3} \sin^2 \theta_W)^2}{1 - \frac{M_Z^2}{m_t^2}} \bigg ]^2 \;
\frac{|\Psi_{00}(0)|^2  }{m^3_{\eta_{gt}}} \; \sin^2 \theta \; 
\\  \nonumber
 \; +  \; \frac{ 8 \pi}{\sqrt{2 \pi}} \;  \bigg [ 1 + (1 - \frac{8}{3} \sin^2 \theta_W)^2 \bigg ] \; 
\frac{1 - \frac{M_Z^4}{4 m_t^4}}{1 - \frac{M_Z^2}{m_t^2}} \; 
   \frac{ a_{Z Z}}{f_{\eta_{gt}}} \;  \frac{\Psi_{00}(0)  }{\sqrt{m_{\eta_{gt}}}}   \; \sin \theta  \; 
 \\ \nonumber  
 + \;  \frac{ a^2_{Z Z }}{3 \pi} \; \bigg [ 1 - \frac{M_Z^4}{4 m_t^4} \bigg ] \;    \frac{m^2_{\eta_{gt}}}{ f^2_{\eta_{gt}} }
 \bigg \} \; ,
\end{eqnarray}
and
\begin{equation}
\label{3.19}
\Gamma \big ( \eta_{gt}  \rightarrow Z_0 \gamma  \big )  \; \simeq  \;  6 \pi  \;  \alpha^2 \; Q_t^2 \;
\frac{\bigg [1 + (1 - \frac{8}{3} \sin^2 \theta_W)^2 \bigg ]^2}{\sin^{2} \theta_W  \cos^{2} \theta_W} 
\;  m_{\eta_{gt}} \; \bigg [ 1 + \frac{M_Z^2}{4 m_t^2} \bigg ]  \; 
 \frac{|\Psi_{00}(0)|^2  }{m^3_{\eta_{gt}}} \; \sin^2 \theta \; .
\end{equation}
\subsection{$\eta_{tg} \rightarrow W^{+} W^{-} $}
\label{s-3.3}
Finally, we consider the decay into two vector bosons $W^+ W^-$. The $W^+ W^-$ decays are similar to the $ZZ$ decays. There are,
however, some differences. In fact,  the pseudoscalar meson $\eta_{gt}$ is allowed to decay into  $W^+ W^-$ pair by means of
the quark-exchange Feynman diagram depicted in Fig.~\ref{Fig-3}, due to the fact that the $\gamma$- and $Z$-exchange
diagrams do not contribute~\cite{Barger:1987}. The transition amplitude corresponding to the Feynman diagram in 
 Fig.~\ref{Fig-3} is readily evaluated:
\begin{eqnarray}
\label{3.20}
\mathcal{A} \big ( \eta_{gt}  \rightarrow  W^+ W^-  \big )  \, \simeq  \, \frac{  2  \pi \alpha}{\sin^2 \theta_W} \,
|U_{tb}|^2 \,   \sqrt{ \frac{3}{M_{\eta_{gt}}}} \;  \frac{\Psi_{00}(0) \, \sin \theta}{(\frac{P}{2} - k_1)^2 - m^2_b} \; 
 \epsilon_{\mu\nu\rho\sigma}  \; \varepsilon_1^{\mu} \, \varepsilon_2^{\nu} \,
P^{\rho} \,  k_1^{\sigma} \; .
\end{eqnarray}
In Eq.~(\ref{3.20}) we considered only the contribution due to the bottom quarks. Since $m_b \ll m_t$, in the following we
set $m_b \simeq 0$. Moreover, to a good approximation we may assume for the weak charged-current mixing matrix element 
$U_{tb} \simeq 1.0$. To obtain the partial decay width we need to consider also the anomalous transition amplitude  due to
the effective Lagrangian Eq.~(\ref{3.17}). Proceeding as in the previous calculations one finds:
\begin{eqnarray}
\label{3.21}
\Gamma \big ( \eta_{gt}  \rightarrow W^+  W^-  \big )  \; \simeq  \;  \frac{1}{16 \pi} \, \frac{\alpha^2}{\sin^{4} \theta_W} 
\, m_{\eta_{gt}} \; 
\bigg \{96 \, \pi^2   \; \frac{1}{\bigg [ 1 - \frac{M_W^2}{m_t^2} \bigg ]^2}  \;
\frac{|\Psi_{00}(0)|^2  }{m^3_{\eta_{gt}}} \; \sin^2 \theta \; 
\\  \nonumber
 \; +  \; \frac{ 16 \pi}{\sqrt{2 \pi}} \;   
\frac{1 - \frac{M_W^4}{4 m_t^4}}{1 - \frac{M_W^2}{m_t^2}} \; 
   \frac{ a_{W W}}{f_{\eta_{gt}}} \;  \frac{\Psi_{00}(0)  }{\sqrt{m_{\eta_{gt}}}}   \; \sin \theta  \; 
 + \;  \frac{ a^2_{W W }}{3 \pi} \; \bigg [ 1 - \frac{M_W^4}{4 m_t^4} \bigg ] \;    \frac{m^2_{\eta_{gt}}}{ f^2_{\eta_{gt}} }
 \bigg \} \; .
\end{eqnarray}
\begin{figure}
\vspace{-0.5cm}
\begin{center}
\includegraphics[width=1.1\textwidth,clip]{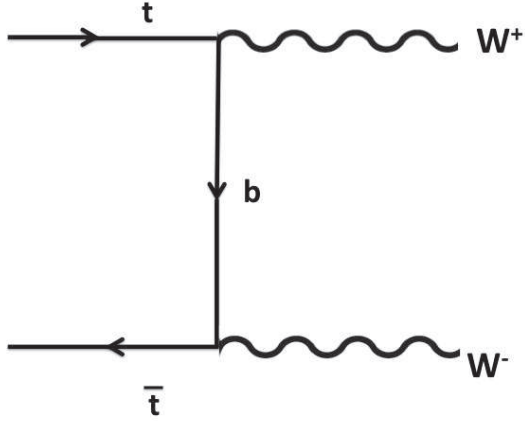}
\caption{\label{Fig-3}  Feynman diagram contributing to the decay of the pseudoscalar $t\bar{t}$ meson into 
 $W^+$ $W^-$. }
\end{center}
\end{figure}
\subsection{$\eta_{tg} \rightarrow \bar{f} f $}
\label{s-3.4}
The pseudoscalar meson $\eta_{gt}$ can decay into a fermion-antifermion pairs via the exchange of virtual $\gamma$ or $Z$.
However, by charge conjugation the  $\gamma$-exchange contribution vanishes, while the $Z$-exchange term contributes
only through the axial-vector coupling. As a consequence the transition amplitude turns out to depend on the fermion mass.
Therefore, the dominant modes are the $b \bar{b}$ and $\tau^+ \tau^-$ decays. The relevant calculations has been 
already presented in the literature (see, for instance Ref.~\cite{Barger:1987} and references therein). For completeness,
we report here the  partial  decay widths:
\begin{equation}
\label{3.22}
\Gamma \big ( \eta_{gt}  \rightarrow b  \bar{b} \big )  \; \simeq  \;  \frac{9 \pi}{8}  \;  \frac{\alpha^2}{\sin^4 \theta_W  \cos^4 \theta_W} 
\; |\Psi_{00}(0)|^2  \; \sin^2 \theta \; \frac{m_b^2}{M^4_Z} \;  ,
\end{equation}
\begin{equation}
\label{3.23}
\Gamma \big ( \eta_{gt}  \rightarrow \tau^+ \tau^-  \big )  \;  \; \simeq  \;  \frac{3 \pi}{8}  \;  \frac{\alpha^2}{\sin^4 \theta_W  \cos^4 \theta_W} 
\; |\Psi_{00}(0)|^2  \; \sin^2 \theta \; \frac{m_{\tau}^2}{M^4_Z} \;  .
\end{equation}
\subsection{Partial Widths and Branching Ratios }
\label{s-3.5}
In the previous Sections we have seen that the pseudoscalar meson may mimic the decays expected for the Standard Model
Higgs boson. As concern the coupling of the pseudoscalar meson $\eta_{gt}$ to the  gauge vector bosons, it turns out 
that we can tune the parameters of our model to  match as much as possible the branching ratios of the Higgs boson at $125 \, GeV$ 
in  the decays into two vector gauge bosons. To estimate the partial decay widths we use the following numerical values:
\begin{equation}
\label{3.24}
  \alpha_s(m_{\eta_{gt}}) \; \simeq \; 0.10 \; \; , \; \;      \alpha(m_{\eta_{gt}})  \;  \simeq   \; \frac{1}{127} \; \;  , \; \; 
  \sin^2 \theta_W \; \simeq \; 0.23 \; \; , \; \; \sin^2 \theta \; \simeq \; 0.237 \; .
\end{equation}
It turns out that the main decay mode of the pseudoscalar  $\eta_{gt}$ is the hadronic decay into two gluons.
From Eq.~(\ref{3.13}) one can easily check that the hadronic decay width is comparable to the total decay width of 
the Standard Model Higgs boson with the same mass. For concreteness, we may tune the parameter $a_{gg}$ such that
the total width of the pseudoscalar meson $\eta_{gt}$ is:
\begin{equation}
\label{3.25}
 \Gamma_{\eta_{gt} } \; \simeq \; 1.0 \;  MeV \; .
\end{equation}
As concern the other parameter of our phenomenological model, we try to fix the values such as to follow as close
as possible the branching ratios of the Standard Model Higgs boson. We found that there are several possibilities.
For illustrative purposes we chosen:
\begin{equation}
\label{3.26}
 a_{gg} \; \simeq \; 0.0170 \; \; , \; \;      a_{\gamma \gamma} \;  \simeq   \; 0.0145 \; \;  , \; \; 
 a_{WW} \; \simeq \; - 0.00055 \; \; , \; \; a_{ZZ} \; \simeq \; - 0.0028 \; .
\end{equation}
%
%
\begin{table}[htb]
\begin{center}
\begin{tabular}{|c|c|c|c|c|}
\multicolumn{4}{c} {} \\
\hline
$Decay \; Channel$  &  $\Gamma(\eta_{gt}) $ & $Br(\eta_{gt})$   & $Br(H)$  \\
\hline
$Z Z $  &   0.0287  MeV  &   0.0287 &   0.0266   \\
\hline
$W W $ &   0.228 \, MeV  &   0.228  &   0.216 \\
\hline
$\gamma \gamma$  &   2.198 \, KeV &   0.0022 &   0.0023  \\
\hline
 $Z  \gamma$ &   3.471 \, KeV &   0.0035 &   0.0016  \\
 \hline
 $b \bar{b} $ &   3.081 10$^{-4}$  MeV & 3.081 10$^{-4}$ &   0.577  \\
\hline
$\tau^+ \tau^-$ &   1.302 10$^{-5}$  MeV &1.302 10$^{-5}$ &   0.064  \\
\hline
$ g g $ &  0.747 \;  MeV &   0.747 &   0.086  \\
\hline
\end{tabular}
\end{center}
\caption{Summary of partial decay widths and  branching ratios of the pseudoscalar meson $\eta_{gt}$.
 For comparison we display  the Standard Model Higgs boson branching ratios~\cite{Heinemeyer:2013}.} 
\label{Table-1}
\end{table}
%
In Table~\ref{Table-1} we display the resulting partial decay widths and branching ratios of the $\eta_{gt}$ pseudoscalar
resonance.  For comparison we, also, show the branching ratios of the Standard Model Higgs boson.
From Table~\ref{Table-1} we see, then, that our peculiar pseudoscalar 
could mimic the decays of the Standard Model Higgs boson in all channels with the exception of the decay into
two fermions  that turns out to be naturally suppressed.  Note that the main decay modes of the pseudoscalar meson
are the hadronic decays. However, due to the huge QCD background these decays are extremely difficult to 
detect experimentally at LHC.
\section{Cross Section and Interaction Strengths}
\label{s-4}
In order to check that the new LHC resonance is, indeed, the Standard Model Higgs boson 
one may compare to observations  the rate for production of the Standard Model Higgs boson in a given decay channel.
To this end, one introduces the relative signal strength defined as the ratio between the observed
signal rate from fit to data to the expected Standard Model signal rate at the given mass.  
The ATLAS and CMS experiments  have presented the values of the signal strengths  obtained in the LHC Run 1 
operations for various decay channels of the $H$ resonance.  In particular, in Ref.~\cite{ATLAS-CMS:2016} it
is reported the observed values of the  $H$ resonance interaction strengths defined by: 
\begin{equation}
\label{4.1}
\mu_{F}  \; = \; \frac{\sigma_H^{obs} \; Br^{obs}( H \rightarrow F)}{ \sigma_H^{th} \; Br^{th}( H \rightarrow F)}  \;  \; ,
\end{equation}
where $\sigma_H^{th}$ and $Br^{th}( H \rightarrow F)$ are the  Standard Model predictions for the
inclusive production cross section and the relevant branching ratio assuming that the $H$ resonance is
the Higgs boson.  It is useful to introduce the following interaction strength for the $\eta_{gt}$ meson:
\begin{equation}
\label{4.2}
\mu^{\eta_{gt}}_{F}  \; = \; 
\frac{\sigma_{\eta_{gt}}^{th} \; Br^{th}( \eta_{gt} \rightarrow F)}{ \sigma_H^{th} \; Br^{th}( H \rightarrow F)}  \;  \; .
\end{equation}
In fact, if the LHC $H$ resonance turns out to be the  $\eta_{gt}$ meson, then the measured interaction strengths
should satisfy:
\begin{equation}
\label{4.3}
\mu_{F}   \; \simeq \; \mu^{\eta_{gt}}_{F}  \;  \; .
\end{equation}
We see, then, that to compare quantitatively our theoretical proposal with observations it is enough to estimate 
the interaction strengths  $\mu^{\eta_{gt}}_{F} $, Eq.~({\ref{4.2}). In Sect.~\ref{3.5} we already estimated the 
$\eta_{gt}$ meson branching ratios. Therefore, now it is necessary to evaluate the  $\eta_{gt}$  inclusive
production cross section. Since the $\eta_{gt}$  meson is an admixture of the pseudoscalar glueball  $\eta_{g}$
and the pseudoscalar $t \bar{t}$ bound state $\eta_{t}$, the production cross section can be estimated  
from the production cross sections of the relevant hadronic bound states. Unfortunately, the production
mechanisms of a given bound state in hadron collisions at high energies is strongly model dependent.
Moreover, no single hadro-production model  is able to describe all the experimental data
(see, eg,  Ref.~\cite{Brambilla:2011} and references therein). Actually, in several models  to describe
 heavy quarkonium  $q \bar{q}$  bound-state production the inclusive  cross section
 is expected to be a fraction ${\cal{F}}$  of  the cross section of the produced $q  \bar{q}$ pairs. 
 Indeed,  these model are still widely used as simulation benchmark since,  once the fractions are determined, it has a full predicting power about cross sections. In this case,  the coupling of a specific bound state to the $q\bar{q}$  pair is directly determined by the appropriate wave function which includes all relevant quantum number projections in conformity with the spin, angular momentum, 
 charge conjugation and the color singlet nature of the bound state considered.
We already observed that, since $m_t \gg m_{\eta_g}$, the $t \bar{t}$ bound-state wave function  overwhelms
the pseudoscalar glueball wave function. Therefore, naively, one expects that the main production mechanism of the 
$\eta_{gt}$ meson is through   $t \bar{t}$ pairs. However, one should keep in mind that the production cross
section of two gluons is expected to exceed the $t \bar{t}$ cross section by orders of magnitude. So that, in general,
both mechanisms should contribute to the associated production of the  $\eta_{gt}$ pseudoscalar meson.
Nevertheless, we shall assume that the main production mechanism is due to the inclusive
$t \bar{t}$ cross section. This is certainly  a rather crude procedure, yet  the resulting values for the inclusive 
production cross sections should be reliable enough to our purposes.  According to our previous discussion, we assume that:
\begin{equation}
\label{4.4}
\sigma^{th}_{\eta_{gt}}   \; \simeq \; {\cal{F}}  \; \sigma^{th}_{t \bar{t}}  \;  \; .
\end{equation}
For the theoretical $t \bar{t}$ pair production cross section we use the top-quark-pair cross
section available in Ref.~\cite{TtbarNNLO}. \\
 In Table~\ref{Table-2} we report the inclusive $t \bar{t}$ cross sections
at $\sqrt{s} = 7, 8, 13 \, TeV$. To determine the parameter  ${\cal{F}}$  we impose that at   $\sqrt{s} = 8  \, TeV$:
\begin{equation}
\label{4.5}
\sigma^{th}_{\eta_{gt}} (\sqrt{s} = 8 \, TeV)  \; \simeq \; \sigma^{th}_{H} (\sqrt{s} = 8 \, TeV)  \;  \; ,
\end{equation}
where  $\sigma^{th}_{H}$ is the theoretical  inclusive production cross section of the Standard Model Higgs boson,
also displayed  in Table~\ref{Table-2}. 
%
%
\begin{table}[t]
\begin{center}
\begin{tabular}{|c|c|c|c|c|}
\multicolumn{5}{c} {} \\
\hline
$\sqrt{s} \, (TeV) $  &  $\sigma^{th}_{t \bar{t}}(pb) $ & $\sigma^{th}_{\eta_{gt}}(pb)$   & $\sigma^{th}_{H}(pb)$ 
& $\sigma^{obs}_{H}(pb)$ \\
\hline
$7 $  &   $173.60^{+11.25}_{-11.79}$   &   $15.20^{+0.98}_{-1.03}$  &  $17.4 \, \pm \, 1.6$   &  $22.1^{+7.5}_{-5.9}$  \\
\hline
$8 $ &   $247.74^{+16.02}_{-15.94}$   &   $21.69^{+1.40}_{-1.40}$  &  $22.3 \, \pm \, 2.0$   &  $27.7^{+3.8}_{-3.6}$   \\
\hline
$13$  &   $815.96^{+45.51}_{-49.82}$   &   $71.44^{+3.98}_{-4.36}$  &  $55.5^{+2.4}_{-3.4}$   &  $59.0^{+10.7}_{-9.8}$   \\
\hline
\end{tabular}
\end{center}
\caption{Summary of the  production inclusive cross sections at $\sqrt{s} = 7, 8, 13 \; TeV$.} 
\label{Table-2}
\end{table}
%
 At $\sqrt{s}= 7, 8  \,  TeV$  
 the theoretical inclusive  Higgs boson production  cross sections have been  taken from Table~1 in Ref.~\cite{ATLAS:2016},
while at $\sqrt{s}= 13 \, TeV$ the theoretical Higgs production cross section has been taken from  
Table~8 in Ref.~\cite{ATLAS:2016ab}.  \\
From Table~\ref{Table-2} and Eq.~(\ref{4.5}) we readily obtain:
\begin{equation}
\label{4.6}
 {\cal{F}}  \; \simeq \; 0.0876  \;  \; .
\end{equation}
After that, our estimate for the inclusive production cross section of the  $\eta_{gt}$ pseudoscalar resonance 
is based on  Eqs.~(\ref{4.4}) and (\ref{4.6}).   For comparison, in  Table~\ref{Table-2} we report the 
observed production cross section of the $H$ resonance.
At $\sqrt{s}= 7, 8  \,  TeV$  the observed production cross section of the $H$ resonance is taken  from Table~5 in
 Ref.~\cite{ATLAS:2016}. At $\sqrt{s}= 13 \, TeV$   the observed $H$  boson  production cross section is based on 
the combined measurements  using more than $13.3 \, fb^{-1}$ of proton-proton collision data recorded by
the ATLAS experiment at the LHC at $\sqrt{s} = 13 \, TeV$. The combination is based on the analyses of the $H$ resonance decays 
into $\gamma \gamma$ and $Z Z \rightarrow 4 \ell$~\cite{ATLAS:2016ab}.
Remarkably,  we see that the our estimate of the $\eta_{gt}$ production cross sections are comparable to 
the theoretical Higgs production cross sections. Moreover, both cross sections are in reasonable agreement with
the experimental  $H$  boson  production cross sections. \\
%
\begin{table}[t]
\begin{center}
\begin{tabular}{|c|c|c|c|c|}
\multicolumn{4}{c} {} \\
\hline
$Decay \; Channel$  & ATLAS & CMS  & $\eta_{gt}$  \\
\hline
$\mu_{Z Z} $  &   $1.52^{+0.40}_{-0.34}$ &   $1.04^{+0.32}_{-0.26}$ &   $1.05 \, \pm \, 0.10$  \\
\hline
$\mu_{W W} $ &   $1.22^{+0.25}_{-0.21}$  & $0.90^{+0.25}_{-0.21}$ &    $1.03 \, \pm \, 0.11$ \\
\hline
$\mu_{\gamma \gamma}$  &   $1.14^{+0.27}_{-0.25}$ &    $1.11^{+0.25}_{-0.23}$ &   $0.94 \, \pm \, 0.10$  \\
\hline
 $\mu_{Z  \gamma}$ &   $2.7^{+4.5}_{-4.3}$ &    $2.7^{+4.5}_{-4.3}$ &    $2.18 \, \pm \, 0.24$  \\
 \hline 
$\mu_{b \bar{b}} $ &    $0.62^{+0.37}_{-0.37}$ &  $0.81^{+0.45}_{-0.43}$ &    $0.00052  \, \pm \, 0.00006$  \\
\hline
$\mu_{\tau^+ \tau^-}$ &    $1.41^{+0.40}_{-0.38}$ & $0.88^{+0.30}_{-0.28}$ &   $ 0.00020 \, \pm \, 0.00002 $   \\
\hline
\end{tabular}
\end{center}
\caption{The measurements of the $H$ resonance interaction strengths performed by  ATLAS and CMS
experiments on data collected during the LHC Run 1.
The data  have been taken from Table~5 in Ref.~\cite{ATLAS-CMS:2016}.
Estimates of the interaction strengths for the $\eta_{gt}$ resonance as given by Eq.~(\ref{4.2}).} 
\label{Table-3}
\end{table}
%
Having determined the inclusive production cross sections of the pseudoscalar meson $\eta_{gt}$, we may
evaluate the interaction strengths Eq.~(\ref{4.2}).  Our results are displayed in Table~\ref{Table-3} together 
with  the measurements of the $H$ resonance interaction strengths performed by  ATLAS and CMS
experiments on data collected during the LHC Run 1. We see that, within the statistical uncertainties, the
$\eta_{gt}$ interaction strengths are in fair good agreements with observations in the vector boson decay 
channels. On the other hand, for the decays into fermions  , the $\eta_{gt}$ interaction strengths
are vanishingly small. Therefore,  we are suggesting that to identify the new LHC resonance with the Standard Model 
Higgs boson it is of fundamental importance to determine experimentally the coupling to fermions to a high degree
of statistical significance. Fortunately, the Run 2 of the LHC at $\sqrt{s}  = 13 \, TeV$ is now making an even larger sample 
of $H$ boson events available for analysis. Nevertheless, the preliminary data from Run 2 does not  show yet
statistically convincing  evidences of the decay $ H  \rightarrow b \bar{b}$ , that should be   the dominant decay 
mode of the Standard Model Higgs boson.  On the other hand, we already noticed that the both ATLAS and
CMS experiments reported new measurements of the $\mu_{t\bar{t}H}$ interaction strength 
confirming an enhanced  coupling of the $H$ boson to the top quarks. Note that, within our approach  
this enhancement with respect to theoretical expectations simply means that the  production cross section
of the  $\eta_{gt}$ meson associated to a $t \bar{t}$ pair is about a factor two higher with respect to the Standard
Model Higgs boson. In any case, we feel that will be interesting to follow as more data are recorded.
\section{The $H_T$ Higgs Boson}
\label{s-5}
In Ref.~\cite{Cea:2012} it was discussed  the scenario  where the Higgs boson without self-interaction  could coexist 
with the spontaneous symmetry breaking of the gauge symmetries.  In that case, as we said in Sect.~\ref{s-1},  
the relation between Higgs mass and the physical Higgs  condensate  is not the same as in perturbation theory leading to the remarkable prediction Eq.~({\ref{1.4}) that implies  a rather massive  Higgs boson, $m_{H_T}  \simeq 750 \,  GeV$. 
 The coupling of the Higgs field to the gauge vector bosons is fixed by the gauge symmetries. Therefore, the coupling of
 the $H_T$ Higgs boson to the  gauge vector bosons  is the same as in perturbation theory notwithstanding the
 non perturbative Higgs condensation driving the spontaneous breaking of the gauge symmetries. 
Given the rather large mass of the $H_T$ Higgs boson, the main decay modes are the decays into two massive
vector bosons (see, e.g., Refs.~\cite{Gunion:1990,Djouadi:2008}):
\begin{equation}
\label{5.1}
\Gamma( H_T \; \rightarrow \; W^+ \, W^-)  \; \simeq  \;  \frac{G_F \, m^3_{H_T}}{8 \pi \sqrt{2}} \;
 \sqrt{1 - \frac{4 m^2_W}{m^2_{H_T}}} \;  \bigg ( 1 - 4 \,  \frac{m^2_W}{m^2_{H_T}} + 12 \, \frac{ m^4_W}{m^4_{H_T}}
 \bigg ) \; 
\end{equation}
and
\begin{equation}
\label{5.2}
\Gamma( H_T \;  \rightarrow \;  Z^0 \, Z^0) \; \simeq \;   \frac{G_F \, m^3_{H_T}}{16 \pi \sqrt{2}} \;
 \sqrt{1 - \frac{4 m^2_Z}{m^2_{H_T}}} \; \bigg ( 1 - 4 \,  \frac{m^2_Z}{m^2_{H_T}} + 12 \, \frac{ m^4_Z}{m^4_{H_T}}
 \bigg )  \; . 
\end{equation}
 On the other hand, it is known that for heavy Higgs the radiative corrections to the decay widths can be safely 
 neglected~\cite{Fleischer:1981,Fleischer:1983,Marciano:1988}. \\
 The couplings of the $H_T$ Higgs boson to the fermions are given by the Yukawa couplings $\lambda_f$. Unfortunately,
 there are not reliable lattice non-perturbative simulations on the continuum limit of the Yukawa couplings. If we follow the
 perturbative approximation, then the fermion Yukawa couplings turn out to be proportional to the fermion mass,
 $\lambda_f = m_f/v$. In that case, for heavy Higgs the only relevant fermion coupling is the top Yukawa coupling 
 $\lambda_t$. On the other hand, we cannot exclude that the couplings of the physical Higgs field  to the fermions
  could be very different from  perturbation theory.  Therefore, to parametrize  our ignorance on the Yukawa couplings
 of the $H_T$ Higgs boson we introduce the parameter:
\begin{equation}
\label{5.3}
\kappa \; = \; \lambda_t^2 \; \frac{v^2}{m_t^2} \; .
\end{equation}
Obviously, in perturbation theory we have $\kappa = 1$. However, we shall also consider the case of $\kappa \simeq 0$
corresponding to strongly suppressed fermion  Yukawa  couplings. \\
The width for the decays of the $H_T$ boson into a $t \bar{t}$ pairs is easily found~\cite{Gunion:1990,Djouadi:2008}:
\begin{equation}
\label{5.4}
\Gamma( H_T \rightarrow \; t \, \bar{t}) \; \simeq \; \kappa \,   \frac{3 \, G_F \, m_{H_T} m^2_t}{4 \pi \, \sqrt{2}} \;
\bigg ( 1 - 4 \,  \frac{m^2_t}{m^2_{H_T}}  \bigg )^{\frac{3}{2}}  \; . 
\end{equation}
So that, to a good approximation, the Higgs total width is given by:
\begin{equation}
\label{5.5}
\Gamma_{H_T}  \; \simeq \; \Gamma( H_T \rightarrow W^+ \, W^-)  \; + \; \Gamma( H_T \rightarrow Z^0 \, Z^0)  \; + \;
 \Gamma( H_T \rightarrow t \, \bar{t})  \; .
\end{equation}
In Fig.~\ref{Fig-4}, left panel, we display  the total width  $\Gamma_{H_T}$, Eq.~(\ref{5.5}), versus the $H_T$ Higgs mass for
both $\kappa = 1$ and $\kappa = 0$. We see that, in the high mass region $m_{H_T} \gtrsim 400 \,  GeV$  the  total width depends
 strongly on the Higgs mass.
\begin{figure}
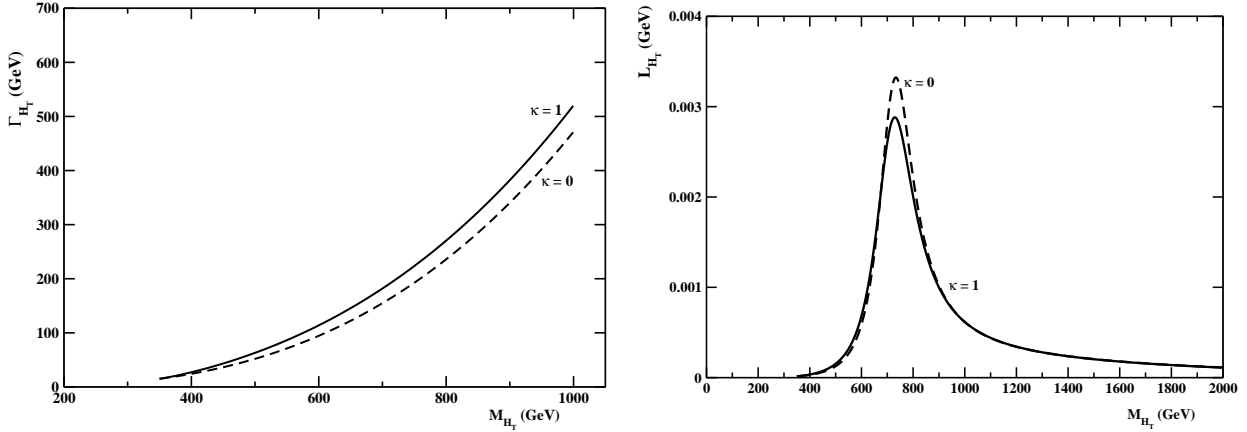

\includegraphics[width=0.5\textwidth,clip]{Fig-4a.eps}
\hspace{0.2 cm}
\includegraphics[width=0.5\textwidth,clip]{Fig-4b.eps}
\caption{\label{Fig-4} The $H_T$ Higgs total width (left panel) and   Lorentzian distribution (right panel)
versus $m_{H_T}$ for $\kappa = 1$ (full line) and $\kappa = 0 $ (dashed line).}
\end{figure}
The main difficulty in the experimental identification of a very heavy  Higgs   resides in the  
large width which makes almost impossible to observe a mass peak.  In fact, the expected  mass spectrum of our heavy Higgs 
 should be proportional to the Lorentzian distribution. Indeed, for a resonance with mass $M$ and total width $\Gamma$
 the Lorentzian distribution is given by:
\begin{equation}
\label{5.6}
L (E) \;  \sim  \;  \; \frac{\Gamma}{(E \; - \; M^2)^2 \; + \;  \Gamma^2/4}  \;  .
\end{equation}
Note that  Eq.~(\ref{5.6}) is the simplest distribution consistent with the Heisenberg uncertainty principle and the finite lifetime
$\tau \simeq 1/\Gamma$.  We obtain, therefore, the following Lorentzian  distribution (see Fig.~\ref{Fig-4}, right panel):
\begin{equation}
\label{5.7}
L_ {H_T} (E) \;  \simeq  \;  \; \frac{1}{ 1.0325 \;  \pi} \; \frac{\frac{\Gamma_{H_T}(E)}{2}}{\big (E \; - \; 750 \; GeV \big )^2 \; + 
\;  \big ( \frac{\Gamma_{H_T}(E)}{2} \big )^2} \;  \;   \;  \; , 
\end{equation}
where $\Gamma_{H_T} ( E)$ is given by Eq.~(\ref{5.5}), and the normalization is  such that:
\begin{equation}
\label{5.8}
\int^{\infty}_{0} \; L_ {H_T} (E)  \;  dE \; \; = \; \; 1 \; \; .
\end{equation}
Note that, in the limit $\Gamma_{H_T} \; \rightarrow \; 0$,  $L_ {H_T} (E)$ reduces to $\delta(E \; - 750 \; GeV)$.  \\
To evaluate the Higgs event production at LHC we need the inclusive Higgs production cross section. As in perturbation
theory, for large Higgs masses the main production processes are by vector-boson fusion and gluon-gluon fusion. 
In fact, the $H_T$ Higgs production cross section by vector-boson fusion is the same as in the perturbative Standard Model calculations.
Moreover, for Higgs mass in the range $700 - 800 \; GeV$ the main production mechanism at LHC is 
expected to be  by the gluon fusion mechanism.  The gluon coupling to the Higgs boson in the Standard Model is 
mediated by triangular loops of top and bottom quarks.  Since in perturbation theory the Yukawa couplings 
of the Higgs particle to heavy quarks grows with quark mass, thus balancing the decrease of the triangle amplitude, 
the effective gluon coupling  approaches a non-zero value for large loop-quark masses. This means that for
heavy Higgs the gluon fusion inclusive cross section is almost completely  determined by the top quarks, even though
it is interesting to stress that  for large Higgs masses the vector-boson fusion mechanism becomes competitive to 
the gluon fusion Higgs production.
According to our approximations the total inclusive cross section for the production of the $H_T$ Higgs boson
can be written as:
\begin{equation}
\label{5.9}
\sigma(p \; p \;  \rightarrow \; H_T \; + \; X) \; \simeq \;    \sigma_{VV}(p \; p \;  \rightarrow \; H_T \; + \; X)
\; + \; \kappa \; \sigma_{gg}(p \; p \;  \rightarrow \; H_T \; + \; X)  \; ,
\end{equation}
where  $\sigma_{VV}$ and   $\sigma_{gg}$ are the vector-boson fusion and gluon-gluon fusion inclusive cross
sections respectively. \\
\begin{figure}
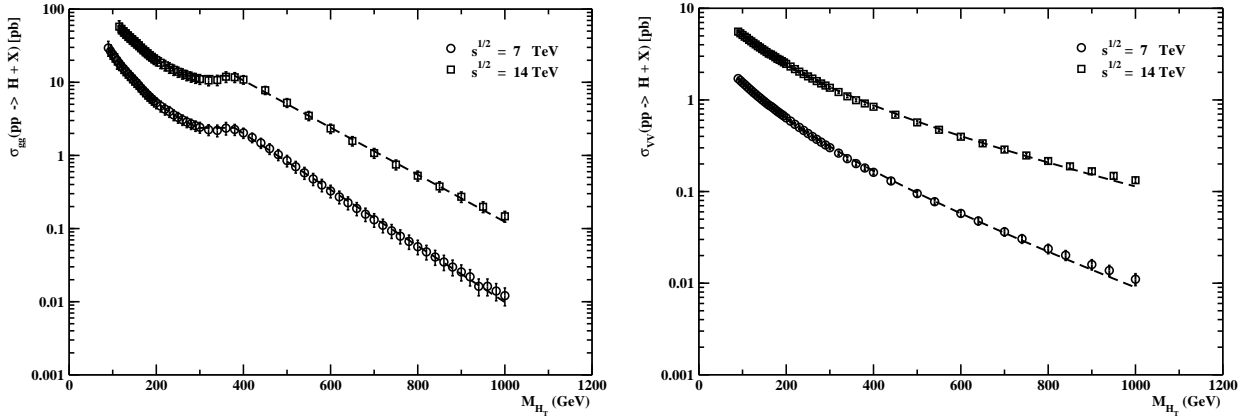

\includegraphics[width=0.5\textwidth,clip]{Fig-5a.eps}
\hspace{0.2 cm}
\includegraphics[width=0.5\textwidth,clip]{Fig-5b.eps}
\caption{\label{Fig-5} The inclusive Higgs production gluon-gluon fusion (left) and vector-boson fusion cross sections (right).
Open circles and squares correspond to the cross sections at $\sqrt{s} = 7 , 14 \, TeV$ respectively.
The data have been taken from  Ref.~\cite{Dittmaier:2011}. The dashed lines are the fits of the data
to our parametrization Eqs.~(\ref{5.10}) and (\ref{5.12}).}
\end{figure}
In Ref.~\cite{Dittmaier:2011} it is presented the calculations of the cross sections computed at next-to-next-to-leading 
and next-to-leading order  for each of the production modes at $\sqrt{s} = 7, 8, 14 \, TeV$. In Fig.~\ref{Fig-5} we show 
the dependence on the Higgs mass of the perturbative  gluon-gluon (left panel) and vector-boson fusion (right panel) 
cross sections at $\sqrt{s} = 7 , 14 \,  TeV$  as reported in  Ref.~\cite{Dittmaier:2011}.   
As concern the gluon-gluon fusion cross section at $\sqrt{s} = 14 \, TeV$   we found that  this cross section can be
 parametrized as:
\begin{equation}
\label{5.10}
 \sigma_{gg}(p \; p \;  \rightarrow \; H_T \; + \; X)  \; \simeq \;  
 \left\{ \begin{array}{ll}
 \;  \left (  \frac{ a_1}{ M_{H_T}} 
 \; + \; a_2 \; M_{H_T}^3  \right )  \;  \exp (-  a_3 M_{H_T})  \; \; &  M_{H_T}  \; \leq \; 273  \; GeV 
  \\
 \; \; \;  \; a_4  \;  & 273 \; GeV    \leq  M_{H_T}   \leq  380 \; GeV
  \\
 \; \;  \; \;a_4 \;  \exp \big [ - a_5 ( M_{H_T} - 380 \; GeV) \big ]  \; \; &  380  \; GeV \; \leq \; M_{H_T}
\end{array}
    \right.
\end{equation}
where $M_{H_T} $  is expressed in  GeV.  In fact we fitted Eq.~(\ref{5.10}) to the data (see the dashed line in 
Fig.~\ref{Fig-5}, left panel) and obtained:
\begin{eqnarray}
\label{5.11}
a_1 \simeq 3.55 \, 10^4 \; pb \, GeV \;  \; , \; \;  a_2 \simeq 3.16 \, 10^{-5} \; pb \, GeV^{-3} \; , \;  
\nonumber \\
a_3 \simeq 1.52 \, 10^{-2} \;  GeV^{-1} \; , \;  \; a_4 \simeq 12.27 \;\, pb  \; , \hspace{2.0 cm}
\\ \nonumber
a_5 \simeq 7.42 \, 10^{-3}  \; GeV^{-1}  \; . \hspace{5.0 cm}
\end{eqnarray}
Likewise,  we parametrized the  dependence of the vector-boson fusion cross section as:
\begin{equation}
\label{5.12}
 \sigma_{VV}(p \; p \;  \rightarrow \; H_T \; + \; X) \; \simeq \;    \bigg ( b_1 \; + \;  \frac{ b_2}{ M_{H_T}} 
 \; + \; \frac{b_3}{ M_{H_T}^2}  \bigg )  \;  \exp (-  b_4 \;  M_{H_T} )   \; ,
\end{equation}
and obtained for the best fit (see the dashed line in  Fig.~\ref{Fig-5}, right panel) :
\begin{eqnarray}
\label{5.13}
b_1 \simeq 3.08 \, 10^{-6}  \; pb   \;  \; , \; \;  b_2 \simeq 7.95 \, 10^{2} \; pb \, GeV \; , \hspace{1.15 cm}
 \nonumber \\
b_3 \simeq - 1.64 \, 10^{4}  \; pb \,  GeV^{2}  \; \;  , \;  b_4 \simeq  1.92 \, 10^{-3} \; GeV^{-1} \; . 
\end{eqnarray}
We have explicitly checked that the same parameterizations are  valid also at $\sqrt{s} = 7 , 8 \, TeV$ ( see Fig.~\ref{Fig-5}).
\subsection{$H_T$ Decay Channels }
\label{s-5.1}
We have seen that the main decays of the heavy $H_T$ Higgs boson are the decays into two massive vector boson
and $t \bar{t}$ pairs, if $\kappa = 1$. Since the search for the decays of the $H_T$ Higgs boson into
pairs of top quarks is very complicated because of the large QCD background, we will focus on the decays
into two massive vector bosons~\footnote{It is worthwhile to stress that the $H_T$ decay to $\gamma \gamma$
do not proceed directly, but, to a fair good approximation, via longitudinal $W$ virtual states. Therefore the relevant
branching ratio turns out be strongly suppressed, i.e.
$Br( H_T \rightarrow \gamma \gamma) \sim 10^{-6}$~\cite{Ellis:1976,Shifman:1979}.}. 
\\
To compare the invariant mass spectrum of our $H_T$ Higgs with the experimental data, we observe that:
\begin{equation}
\label{5.14}
 N_{H_T} (E_{1},E_{2} )  \; \simeq \; {\cal{L}} \;  \int^{E_2}_{E_1} \; Br(E) \;  \varepsilon(E) \; \sigma(p \; p \;  
 \rightarrow \; H_T \; + \; X)  \; L_ {H_T} (E)  \;  dE  
 \;  \; , 
\end{equation}
where  $N_ {H_T}$ is the number of Higgs events in the energy interval $E_1,E_2$, corresponding to an integrated luminosity 
${\cal{L}}$, in the given channel with branching ratio $Br(E)$.   The parameter $ \varepsilon(E)$  accounts for  the efficiency 
of trigger, acceptance of the detectors, the kinematic selections, and so on.  Thus, in general
 $ \varepsilon(E)$ depends on the energy, the selected channel and the detector.   For illustrative purposes,
 we consider the decay  channels  $H  \rightarrow WW  \rightarrow \ell \nu q q$ and $H  \rightarrow ZZ   \rightarrow \ell \ell q q$.
As concern the branching ratios Br(E), we can write:
\begin{equation}
\label{5.16}
  \begin{array}{ll}
Br(H_T  \rightarrow WW  \rightarrow \ell \nu q q ) \;  \simeq  \; Br(H_T  \rightarrow WW  ) \;
\times  \; Br( WW  \rightarrow \ell \nu q q ) 
  \\
Br(H_T  \rightarrow ZZ   \rightarrow \ell \ell q q )  \;  \simeq  \; Br(H_T  \rightarrow ZZ  )  \;
\times  \; Br( ZZ \rightarrow \ell \ell q q ) \; ,
\end{array}
\end{equation}
where
\begin{equation}
\label{5.17}
Br(H_T  \rightarrow WW  ) \; = \; \frac{ \Gamma( H_T \rightarrow WW)}{ \Gamma_{H_T}}   \; \; , \; \; 
Br(H_T  \rightarrow ZZ  ) \; = \; \frac{ \Gamma( H_T \rightarrow ZZ)}{ \Gamma_{H_T}}   \; ,
\end{equation}
and the branching ratios for the decays of $W$ and $Z$ bosons into fermions are given by the Standard Model 
values~\cite{Olive:2014}. \\
In Fig.~\ref{Fig-6}  we display  the distribution of the invariant mass for the
Higgs boson candidates corresponding to the process $H_T \; \rightarrow WW \; \rightarrow \ell \nu qq$ (left) and
 the distribution of the invariant mass for the process $H_T \; \rightarrow ZZ \; \rightarrow \ell \ell qq$ (right).
 The distributions have been obtained using Eq.~(\ref{5.14}) with an integrated luminosity  ${\cal{L}} = 30 \,  fb^{-1}$
 for both $\kappa=1$ and $\kappa = 0$.  We used the Higgs inclusive cross section at $\sqrt{s} = 14 \, TeV$ scaled by
 about $10 \%$ as our estimate of the cross section at $\sqrt{s} = 13 \, TeV$. 
 From Fig.~\ref{Fig-6}  we see that the search strategy is to find some structures
 in the invariant mass spectrum. However, one should keep in mind that such strategy is severely hampered by the
 presence of the so-called irreducible background. The main backgrounds are given by the production of $W + jet$  and   $Z + jet$ which, however,  should be suppressed for large invariant mass.  Moreover, the presence of charged lepton should allows to obtain a good 
rejection of the QCD processes.  \\
\begin{figure}
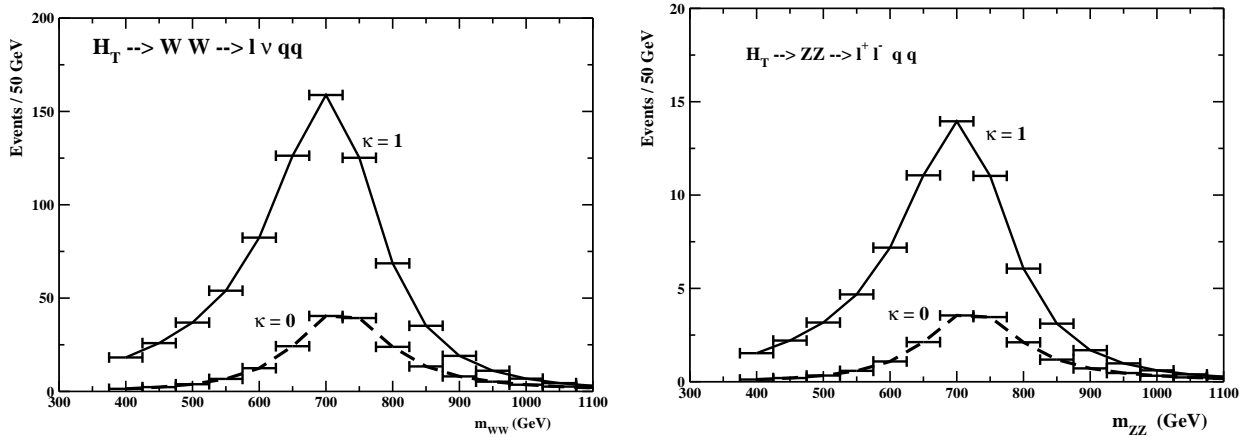

\includegraphics[width=0.5\textwidth,clip]{Fig-6a.eps}
\hspace{0.2 cm}
\includegraphics[width=0.5\textwidth,clip]{Fig-6b.eps}
\caption{\label{Fig-6} (Left) Distribution of the invariant mass $m_{W W}$ for the  process 
$H_T \; \rightarrow WW \; \rightarrow \ell \nu qq$  ($\ell = e, \mu$) corresponding to an integrated luminosity of 30 fb$^{-1}$. 
The  $H_T$ Higgs event distribution (binned in energy intervals of $50 \, GeV$) are evaluated according to Eq.~(\ref{5.14})
 assuming $\varepsilon(E) \simeq  0.20 $, $\kappa = 1$  and $\kappa = 0$.
(Right)  Distribution of the invariant mass $m_{Z Z}$ for the  process 
$H_T \; \rightarrow Z Z \; \rightarrow \ell \ell q q$  ($\ell = e, \mu$) corresponding to an integrated luminosity of $30 fb^{-1}$. }
\end{figure}
It is interesting to mention that preliminary results from the ATLAS collaboration  reported the experimental  search of large
mass resonances  in the decay to $WW$ final states, with one $W$ boson decaying  into an electron or a muon and a neutrino
while the other $W$ boson decaying hadronically~\cite{ATLAS:2016c}. The data correspond to an integrated luminosity
 of $13.2 \,  fb^{-1}$ collected by the ATLAS detector at $\sqrt{s} = 13 \, TeV$.  The data hint to an excess in the invariant mass
 distribution around $700 \, GeV$ that, however, is not yet statistical significant. Likewise, in Ref.~\cite{ATLAS:2016d} 
 it is reported the searches for heavy resonance decaying into $ZZ$ pair using collision data at $\sqrt{s} = 13 \, TeV$
at the Large Hadron Collider corresponding to an integrated luminosity   of $13.2 \, fb^{-1}$ recorded with the ATLAS
detector in 2015 and 2016. In the final states in which one $Z$ boson decays to a pair of light charged leptons (electrons
or muons) and the other $Z$ boson decays into hadrons there is some excess in the invariant mass distribution near
$700 \, GeV$ for the vector-boson fusion production processes. However, it should be mentioned that the eventual
signal is  not statistically significant due to the low integrated luminosity.
\subsection{The Golden Channel: Comparison with the LHC Data}
\label{s-5.2}
The decay channels  $H_T  \rightarrow ZZ  \rightarrow \ell \ell  \ell \ell $ (the golden channel) have very low branching ratios.
Nevertheless, the presence of leptons allows to efficiently reduce the huge  background due mainly to diboson production. 
In this Section we attempt a quantitative comparison of our theoretical expectations with the experimental data
from LHC Run 2. To increase the integrated luminosity we decided to combine the experimental data from
both ATLAS~\cite{ATLAS:2016e}  and CMS~\cite{CMS:2016c} experiments.    In Fig.~\ref{Fig-7} we show 
the invariant mass distribution for the golden channel obtained by combining the data from the ATLAS experiment
with an integrated luminosity of   $14.8 \, fb^{-1}$~\cite{ATLAS:2016e} and the CMS experiment with
an integrated luminosity of   $12.9 \, fb^{-1}$~\cite{CMS:2016c}.  From our estimate of the background 
(dotted line in Fig.~\ref{Fig-7}) we see that, indeed, in the high invariant mass region   $m_{ZZ}  \gtrsim 600 \,  GeV$, 
the background is strongly suppressed.  To compare  with our theoretical expectations, we displayed 
in  Fig.~\ref{Fig-7} the invariant mass distribution corresponding to the signal plus the background (full line).
Due to the rather low integrated luminosity, we see that the signal manifest itself
as an approximate plateaux in the   invariant mass interval   $600 \, GeV \lesssim m_{ZZ}   \lesssim  740 \,  GeV$
over  a smoothly decreasing background. Actually, in this region it seems that  there is a moderate excess of signal
over the expected background that seems to  compare quite well with our theoretical prediction. 
To be qualitative, we estimated the total number of events in the invariant mass interval  
$600 \, GeV \lesssim m_{ZZ}   \lesssim  740 \,  GeV$.  We found that  the expected background event was $N_{back} = 8.8$,
while our  theoretical estimate of the signal  was $N^{th}_{sign} = 6.5$.
The observed events in the given interval was $N_{obs} = 15.0^{+5.5}_{-3.4}$ where, 
to be conservative, the quoted errors have been obtained by adding in quadrature the experimental errors.
So that there is an excess over the expected background with a statistical significance of about $2 \, \sigma$ that
is in fair agreement  with the theoretical expectations.
Therefore, our $H_T$ Higgs event distribution in the golden channel is not in contradiction with the experimental data, even though 
we cannot exclude that the data are compatible with the background-only hypothesis. In fact, we estimated that to meet
to  the $5 \, \sigma$ criterion in High Energy Physics it will be necessary to collect an integrated luminosity 
${\cal{L}} \sim 10^{2} \,  fb^{-1}$.
\begin{figure}
\centering
\includegraphics[width=0.67\textwidth]{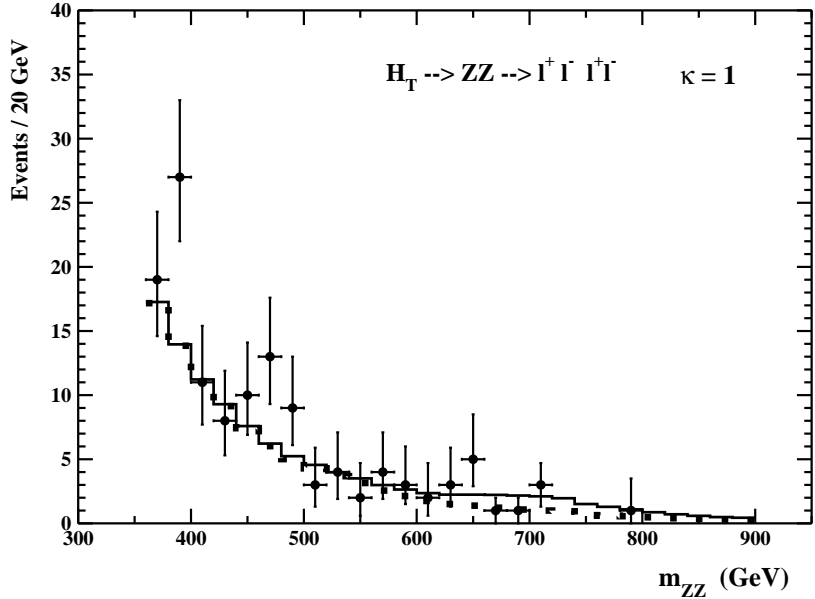}
\caption{\label{Fig-7} Comparison to the LHC data of the distribution of the invariant mass $m_{Z Z}$ for the  process
 $H_T \; \rightarrow ZZ \; \rightarrow \ell \ell \ell \ell$   ($\ell = e, \mu$) corresponding to an integrated luminosity 
 of ${\cal{L}} \simeq 27.7 \,  fb^{-1}$    binned in energy intervals of $20 \, GeV$ and assuming  $\kappa = 1$. 
 The data have been obtained combining
the Run 2 LHC data at $\sqrt{s} = 13 \, TeV$ from ATLAS~\cite{ATLAS:2016e}  and CMS~\cite{CMS:2016c}
with integrated luminosity   ${\cal{L}} = 14.8 \,  fb^{-1}$ and $12.9 \,  fb^{-1}$ respectively. The dotted line
is our estimate of the background, the continuum line is the background plus signal histogram assuming 
$\varepsilon(E) \simeq  0.48$ and $\kappa = 1$.}
\end{figure}
\section{Conclusion}
\label{s-6}
It is widely believed that the new LHC resonance at $125 \, GeV$, denoted  by $H$, is the Standard Model Higgs boson. 
The main motivation for this identification resides in the observed decays of the $H$ resonance  into vector bosons that agree
to a high level of statistical significance with the expected rates of the Standard Model Higgs boson.
To firmly establish that the $H$ resonance is, indeed, the Standard Model Higgs boson, it is necessary to demonstrate
the direct coupling to fermions. If the $H$ resonance is identified with the Higgs boson, then the largest branching
 ratio should be in the decays into bottom-antibottom pair. However, the data collected in the LHC Run 1 by both ATLAS and CMS 
 experiments display a puzzling deficit in this channel. Moreover, the preliminary data at $\sqrt{s} = 13 \, TeV$ still do not show a clear
evidence for the decays in $b \bar{b}$ pair. On the other hand, we already pointed out that from  LHC Run 1 and Run 2 data
one infers an enhancement of the $H$ resonance coupling to the top quark.  The forthcoming data from the LHC Run 2
experiments will be of fundamental importance to settle these problems. In the meantime we believe that there are
already compelling reasons to look for different explanations of the $H$ resonance. In the first part of the present paper,
following Ref.~\cite{Cea:2012b}, we advanced the proposal that the $H$ resonance could be a pseudoscalar meson.
We developed a phenomenological approach where the $H$ boson is a coherent superimposition of the low-lying
pseudoscalar glueball and the $t \bar{t}$ bound state. We showed that this peculiar pseudoscalar meson is able
to mimic the decays in vector bosons at the same rate as the Standard Model Higgs boson with the same mass.
We pointed out that, once the decays into two vector bosons cannot be used to identity univocally the Higgs
boson, to unravel the true nature of the $H$ resonance it is necessary to rely on the decays into fermions and 
on the resonance $CP$ assignment where, however, the reached statistical significance is  still below the 
High Energy criterion of $5 \, \sigma$. Finally, we would like to point out that in this paper we attempted an alternative
interpretation of the new LHC resonance entirely  within the Standard Model. However, at present we cannot
exclude different possibilities beyond the Standard Model Physics. \\
In the second part of the present  paper we focussed on the true Higgs boson denoted by  $H_T$.
Stemming from the known triviality problem, i.e. vanishing self-coupling, that affects self-interacting scalar quantum fields
in four space-time dimensions, we evidenced that the Higgs boson condensation triggering the spontaneous
breaking of the local gauge symmetries needs to be dealt with non perturbatively. In this case, from one hand there is
no stability problem for the condensate ground state, on the other hand the Higgs mass is finitely related to the vacuum 
expectation value of the quantum scalar field and, in principle, it can be evaluated from first principles.
In fact, precise non-perturbative numerical simulations indicated that the $H_T$ Higgs boson mass is consistent with
Eq.~(\ref{1.4})~\cite{Cea:2012}  leading to a rather heavy Higgs boson. We critically discussed the couplings of the $H_T$
Higgs boson to the massive vector bosons and to fermions. We estimated the expected production mechanism
and the main decay modes. Finally, we, also, compared our proposal with the recent results in the golden channel from 
ATLAS and CMS collaborations. We found that the available experimental observations are not in contradiction with our scenario. 
We are confident that forthcoming data from LHC Run 2 will add further support to the heavy Higgs proposal.

\end{document}